\documentclass[final,5p,times,onecolumn]{elsarticle}

\usepackage{amssymb}
\usepackage{amsmath}
\usepackage{mathrsfs}
\usepackage{color}
\usepackage{amsthm}
\usepackage{dcolumn}% Align table columns on decimal point
\usepackage{lineno}
\usepackage{cuted}
\usepackage{cuted}

\usepackage{graphicx}
\usepackage{caption}
\journal{Physics Letters A}

\begin{document}

\begin{frontmatter}

\title{Vibrational resonance in coupled self-learning Duffing oscillators and its application in noisy radio frequency signal processing}

\author[label1]{Jianhua Yang\corref{cor1}}
\ead{jianhuayang@cumt.edu.cn}
\cortext[cor1]{Corresponding author}
\address[label1]{Jiangsu Key Laboratory of Mine Mechanical and Electrical Equipment, School of Mechanical and Electrical Engineering, China University of Mining and Technology, Xuzhou, 221116, Jiangsu, People's Republic of China}

\author[label1]{Litai Lou}

\author[label1]{Shangyuan Li}

\author[label2]{Zhongqiu Wang}
\address[label2]{School of Computer Science and Technology, China University of Mining and Technology, Xuzhou 221116, Jiangsu, People's Republic of China}

\author[label3,label4]{Miguel A. F. Sanju\'an}
\address[label3]{Nonlinear Dynamics, Chaos and Complex Systems Group, Departamento de Fisica, Universidad Rey Juan Carlos, Tulipan s/n, Mostoles, 28933, Madrid, Spain}
\address[label4]{Royal Academy of Sciences of Spain, Valverde 22, 28004 Madrid, Spain}

\begin{abstract}
This work presents a new coupled array of frequency-adaptive Duffing oscillators. Based on learning rules, the natural frequency of each oscillator changes with the external excitation to achieve the frequency-adaptive capability in the response. The frequency range of vibrational resonance in the response is greatly extended through the frequency-adaptive learning rule. Moreover, the theoretical condition for vibrational resonance is derived and its validity is verified numerically. The coupled self-learning Duffing oscillators can also perform signal denoising in strong noise environment, and its performance in signal denoising has been verified through processing the simulated signal and the wireless radio frequency signal under two scenarios. The superiority of vibrational resonance to the conventional denosing methods such as wavelet transform and Kalman filter has also been illustrated by experimental radio frequency signal processing. The combination of broadband frequency adaptability and strong noise-reduction capability suggests that these oscillators hold considerable potential for engineering applications.
\end{abstract}
\end{frontmatter}



\section{Introduction}
%\label{sec1}
There are numerous problems in nature and engineering where coupled oscillators serve as essential models for dynamic analysis. Examples include cardiac pacing, biological clocks, neural networks, synchronous firefly flashing, Josephson junction arrays, wave patterns in chemical reactions, drone formations, power grids, microelectromechanical systems, phase-locked loop arrays, clock networks, and clock synchronization in distributed computing, among others. Therefore, an in-depth exploration of the dynamic characteristics of coupled oscillators holds significant theoretical importance and practical engineering value \cite{Ref1, Ref2, Ref3, Ref4, Ref5, Ref6, Ref7, Ref8}.\\
\indent Coupled oscillators exhibit rich dynamical properties, and many researches have focused on their resonance phenomena. Examples include the amplification of a weak signal using the resonance of a $Y$-shaped structure composed of $N+1$ coupled bistable oscillators or a globally coupled network of FitzHugh-Nagumo oscillators \cite{Ref9, Ref10}, the extraction of signal information under noisy backgrounds via stochastic resonance in coupled systems \cite{Ref11, Ref12, Ref13, Ref14, Ref15, Ref16}, and the enhancement of weak characteristic signals through vibrational resonance \cite{Ref17, Ref18, Ref19, Ref20, Ref21, Ref22, Ref23, Ref24, Ref25, Ref26}. However, based on the traditional resonance theoretical framework, the characteristic frequency that causes resonance in the system generally needs to be taken as a small value, which severely limits the signal processing applications based on resonance methods. Moreover, resonance realized based on re-scale or twice sampling methods face the challenge of selecting system parameters. Under large parameter conditions, the system response becomes highly sensitive, and inappropriate parameter choices not only make it difficult to induce resonance but also hinder the practical hardware implementation of signal processing \cite{Ref27, Ref28, Ref29}. Another way to achieve resonance over a wide frequency range is to employ a frequency-adaptive system, in which the natural frequency varies with the external excitation. The effectiveness of frequency-adaptive oscillators in extending the resonance range has been demonstrated in previous studies \cite{Ref30, Ref31}. However, addressing this problem requires the design of suitable frequency-adaptive learning rules, tailored to the specific goals and nonlinear systems under consideration.\\
\indent The idea of vibrational resonance not only has important theoretical significance, but also has potential engineering application values, such as in signal processing \cite{Ref32, Ref33, Ref34}, fault diagnosis \cite{Ref35, Ref35, Ref35, Ref38}, image processing \cite{Ref39, Ref40}, energy harvesting \cite{Ref41, Ref42, Ref43, Ref44}, vibration reduction and isolation \cite{Ref45, Ref46}, and other fields. Inspired by the above researches, this work proposes an array of coupled self-learning Duffing oscillators to induce vibrational resonance in the case of a high-frequency characteristic signal. At the same time, the excellent performance of the proposed system and algorithm in processing noisy signals are presented.\\
\indent The structure of this work is as follows. Section II introduces the model of the coupled oscillators with a learning rule assigned to each oscillator. It then examines the occurrence of vibrational resonance in the high-frequency range under the action of the learning rules and discusses the influence of the signal amplification factor, the coupling strength, the signal amplitude, the learning rate and the number of oscillators on the strength of vibrational resonance. The appropriate number of oscillators required to achieve a satisfactory resonance effect is also determined. In addition, the theoretical condition for vibrational resonance is also derived and verified numerically. Section III illustrates the noise-reduction capability of vibrational resonance in the coupled self-learning oscillators through numerical simulations and the analysis of wireless radio frequency signals. The proposed vibrational resonance algorithm is shown to process weak characteristic signals embedded in strong noise across a broad frequency range while achieving effective noise suppression. Its superiority to the traditional denoise methods such as waveform transform and Kalman filter is also revealed through two experimental scenarios of radio frequency signal processing. Finally, Section IV presents the main conclusions of this work.
\section{Vibrational resonance in coupled self-learning Duffing oscillators}
%\label{subsec1}
\subsection{The coupled self-learning Duffing oscillators}
Under the simultaneous excitation of two harmonic signals, the classical second-order Duffing equation is
\begin{equation}
\left\{ \begin{array}{l}
 \ddot x + 2\zeta \dot x + {\omega _n}^2x + b{x^3} = f(t) \\
 f(t) = A\cos ({\Omega _1}t) + B\cos ({\Omega _2}t) \\
 \end{array} \right.
\end{equation}
In Eq.~(1), $\zeta$ and $\omega _n$ are the damping ratio and the natural frequency of the corresponding linear system, respectively. The parameter $b$ is the nonlinear stiffness coefficient, and $b>0$. The excitation contains the characteristic signal $A\cos ({\Omega _1}t)$ and the auxiliary signal $B\cos ({\Omega _2}t)$, and their parameters satisfy $A \ll 1$ and $\Omega_1 \ll \Omega_2$.\\
\indent Modify the classic second-order Duffing oscillator to a frequency-adaptive system as follows
\begin{equation}
\left\{ \begin{array}{l}
 \ddot x + 2\zeta \dot x + {\omega ^2}x + b{x^3} = \beta f(t) \\
 \dot \omega  = {k_\omega }f(t) \\
 f(t) = A\cos ({\Omega _1}t) + B\cos ({\Omega _2}t) \\
 \end{array} \right.
\end{equation}
In the original system described by Eq.~(1), the natural frequency $\omega_n$ is replaced by $\omega$ in Eq.~(2), which varies with the external excitation. The frequency-adaptive rule endows the system with a self-learning capability, where $k_\omega$ represents the learning rate. In Eq.~(2), $\beta$ denotes the amplification factor applied to the input signals. The amplification factor $\beta$ not only facilitates coordination with the nonlinear system but also promotes the occurrence of vibrational resonance. Moreover, simultaneously amplifying the two excited signals is straightforward to implement in both algorithm and hardware.\\
\indent Further incorporating coupling between two adjacent oscillators, we consider the following unidirectionally coupled array
\begin{equation}
\left\{ \begin{array}{l}
 {{\ddot x}_1} + 2\zeta {{\dot x}_1} + {\omega _1}^2{x_1} + b{x_1}^3 = \beta f(t) \\
 {{\dot \omega }_1} = {k_\omega }f(t) \\
 {{\ddot x}_i} + 2\zeta {{\dot x}_i} + {\omega _i}^2{x_i} + b{x_i}^3 = \beta [f(t) + \epsilon {x_{i - 1}}]\\
 {{\dot \omega }_i} = {k_\omega }\left[ {f(t) + \epsilon {x_{i - 1}}} \right],\quad i = 2,\;3, \cdots ,\;n \\
 f(t) = A\cos ({\Omega _1}t) + B\cos ({\Omega _2}t) \\
 \end{array} \right.
\end{equation}
Herein, $\epsilon>0$ is the coupling strength which makes the former oscillator as the input of its next oscillator. In the designed self-learning coupled oscillators, the output of $\epsilon x_{i-1}$ is input into the $ith$ oscillator together with the excitation signal. The so-called frequency-adaptive system indicates that its natural frequency is related to the excitation of the system directly. For the oscillator $x_i$, the frequency components in the excitation are completely included in $f(t) + \epsilon x_{i-1}$, so the learning rule here is $k_\omega [f(t) + \epsilon x_{i-1}]$. Apparently, the learning rule makes the natural frequency of the oscillator depends on both the output of the forward oscillator and the external excitations. In other words, the expression of learning rules are not only related to external excitations, but also to the output of its forward oscillator. Therefore, this learning rule can be regarded as a modified type of Hebbian learning rule. The original form of Hebbian learning rule is usually used in a system with a limit cycle,such as the Hopf oscillator and the van der Pol oscillator \cite{Ref47}. We cannot ensure the limit cycle existing in each oscillator of Eq.~(3). Hence, it is much more reasonable to use the modified Hebbian learning rules in Eq.~(3). Our inspiration of the learning rules comes but different from another alternative Hebbian learning rule that has been investigated by Perkins in the frequency-adaptive Duffing oscillator \cite{Ref30, Ref31}. To further express the structure of the coupled self-learning Duffing oscillators, we present Fig.~1. This figure not only shows the structure of a single oscillator, but also the connection and coupling relationships between each oscillator.
\begin{figure}%[h]
\includegraphics[width=0.5\textwidth]{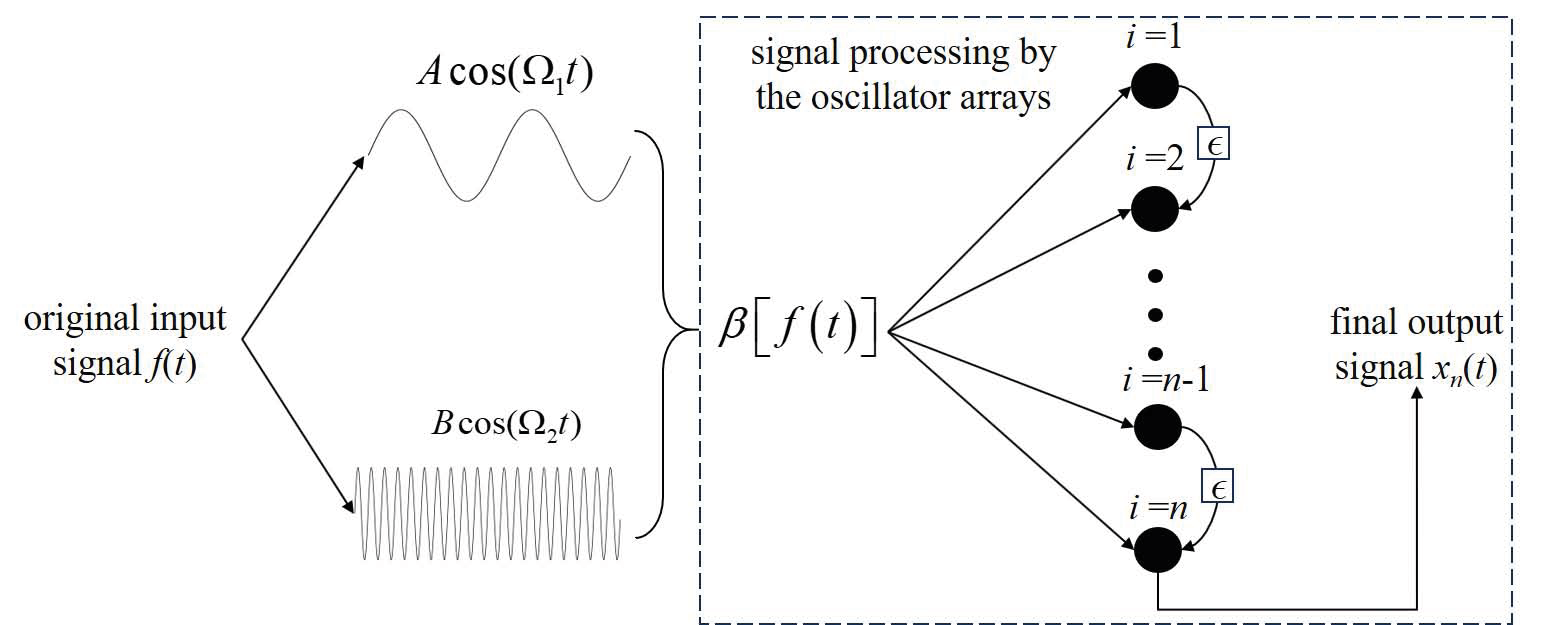}% Here is how to import EPS art
\caption{The structure of the coupled self-learning Duffing oscillators.}
\end{figure}
\subsection{The necessity of the amplification factor $\beta$}
\indent The degree of vibrational resonance is usually measured by the response amplitude at the frequency of the characteristic signal. The response amplitude of the $ith$ oscillator $x_i$ is calculated by
\begin{equation}
Q_i = \frac{{\sqrt {Q_{i\rm{s}}^2 + Q_{i\rm{c}}^2} }}{A},\quad i = 1,\;2, \cdots ,\;n
\end{equation}
where $Q_{i\rm{s}}$ and $Q_{i\rm{c}}$ are the sine and cosine Fourier coefficients of the time series of $x_i$
\begin{equation}
\left\{ \begin{array}{l}
 {Q_{i\rm{s}}} = \frac{2}{{mT}}\int\limits_0^{mT} x_i (t)\sin (\Omega_1 t){\rm{d}}t \\
 {Q_{i\rm{c}}} = \frac{2}{{mT}}\int\limits_0^{mT} x_i (t)\cos (\Omega_1 t){\rm{d}}t \\
 \end{array} \right.
\end{equation}
For numerical simulations, the Fourier coefficients are calculated by
\begin{equation}
\left\{ \begin{array}{l}
 {Q_{i\rm{s}}} = \frac{{2\Delta t}}{{mT}}\sum\limits_{i = 1}^{mT/\Delta t} x_i ({t_i})\sin ({\Omega _1}{t_i}) \\
 {Q_{i\rm{c}}} = \frac{{2\Delta t}}{{mT}}\sum\limits_{i = 1}^{mT/\Delta t} x_i ({t_i})\cos ({\Omega _1}{t_i}) \\
 \end{array} \right.
\end{equation}
where $T=2\pi/\Omega_1$ is the period of the characteristic signal and $m$ is a large enough positive integer. In the following simulations, the time step is $\Delta t=0.001$, the initial conditions are $x(0) = 0$, $\dot x(0) = 0$, and ${\omega _i}(0) = 0, i = 1, 2, \cdots , n$. The total time length is 120 period of the characteristic signal. Then, the time series corresponding to the first 20 cycle length are cut off as the transient response, and the last 100 cycle length, which are $200 \pi/\Omega_1$, is retained as time series for calculating the value of $Q_i$. In other words, we choose $m=100$ in the calculation.\\
\indent If there is no learning rule existing, i.e., if the natural frequency of each oscillator is fixed, vibrational resonance will not occur when the characteristic frequency is high. This fact has been studied in the previous work and will not be repeated here \cite{Ref48}. Figure 2 shows the situation where vibrational resonance occurs in the frequency-adaptive system with learning rules. When the amplification factor $\beta$ is large enough, the response of each oscillator undergoes strong vibrational resonance. As the number of oscillators increases, the degree of vibrational resonance becomes stronger. On the contrary, if the amplification factor $\beta$ is small, vibrational resonance may also exist but the peak value is relatively weak. Therefore, a sufficiently large amplification factor $\beta$ is a condition for strong vibrational resonance to occur, which also determines the size of the resonance peak.
\begin{figure}%[h]
\includegraphics[width=0.5\textwidth]{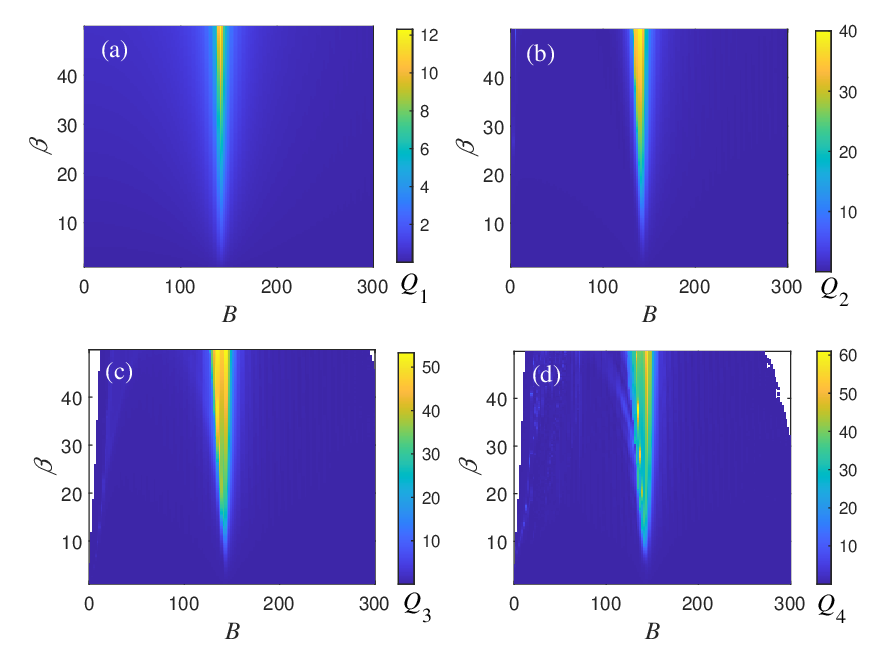}% Here is how to import EPS art
\caption{Influence of the amplification factor $\beta$ on vibrational resonance of Eq.~(3) in the $B-\beta$ plane. The simulation parameters are $\zeta=0.25$, $b=1$, $k_\omega=10$, $A=0.1$, $\Omega_1=10$, $\Omega_2=10\Omega_1$ and $\epsilon=1$. Panels (a)-(d), $Q_i$ versus $B$ and $\beta$, $i=1$, 2, 3, 4 in turn.}
\end{figure}
\subsection{The condition for vibrational resonance of each oscillator}
In Eq.~(3), integrating $\dot \omega_1$ over the time, we have
\begin{equation}
{\omega _1} = \frac{{A{k_\omega }}}{{{\Omega _1}}}\sin ({\Omega _1}t) + \frac{{B{k_\omega }}}{{{\Omega _2}}}\sin ({\Omega _2}t)
\end{equation}
and
\begin{equation}
{\omega _1}^2 = \frac{{{k_\omega }^2}}{2}\left( {\frac{{{A^2}}}{{{\Omega _1}^2}} + \frac{{{B^2}}}{{{\Omega _2}^2}}} \right) - \frac{{{k_\omega }^2{A^2}}}{{2{\Omega _1}^2}}\cos (2{\Omega _1}t) - \frac{{{k_\omega }^2{B^2}}}{{2{\Omega _2}^2}}\cos (2{\Omega _2}t)
\end{equation}
Then, the govern equation of Eq.~(3) for the first oscillator is
\begin{equation}
\begin{array}{l}
 {{\ddot x}_1} + 2\zeta {{\dot x}_1} + \frac{{{k_\omega }^2}}{2}\left( {\frac{{{A^2}}}{{{\Omega _1}^2}} + \frac{{{B^2}}}{{{\Omega _2}^2}}} \right){x_1} - \frac{{{k_\omega }^2{A^2}}}{{2{\Omega _1}^2}}\cos (2{\Omega _1}t){x_1} \\
  - \frac{{{k_\omega }^2{B^2}}}{{2{\Omega _2}^2}}\cos (2{\Omega _2}t){x_1} + b{x_1}^3 = \beta [A\cos ({\Omega _1}t) + B\cos ({\Omega _2}t)] \\
 \end{array}
\end{equation}
The natural frequency of Eq.~(9) can be approximated as
\begin{equation}
{\omega _{n1}} = \sqrt {\frac{{{k_\omega }^2}}{2}\left( {\frac{{{A^2}}}{{{\Omega _1}^2}} + \frac{{{B^2}}}{{{\Omega _2}^2}}} \right)}
\end{equation}
The vibrational resonance occurs when the condition ${\omega _{n1}} = {\Omega _1}$ is satisfied. As a result, we have an approximated vibrational resonance condition for the first oscillator is
\begin{equation}
{\Omega _1} = \sqrt {\frac{{{k_\omega }^2}}{2}\left( {\frac{{{A^2}}}{{{\Omega _1}^2}} + \frac{{{B^2}}}{{{\Omega _2}^2}}} \right)}
\end{equation}
\\
\indent As for the $ith$ ($i>1$) oscillator, due to the main frequency components are $\Omega_1$ and $\Omega_2$ in the majority of cases, we make an approximation for the expression
\begin{equation}
{\dot \omega _i} = {k_\omega }[{A_i}\cos ({\Omega _1}t + {\phi _i}) + {B_i}\cos ({\Omega _2}t + {\varphi _i})],\quad i = 2,3, \cdots ,n,
\end{equation}
where $A_i$, $B_i$, $\phi_i$ and $\varphi_i$ are generated by the combination of the excitation $f(t)$ and the response $\epsilon {x_{i - 1}}$. Integrating $\dot \omega _i$ over the time, we get
\begin{equation}
{\omega _i} = \frac{{{A_i}{k_\omega }}}{{{\Omega _1}}}\sin ({\Omega _1}t + {\phi _i}) + \frac{{{B_i}{k_\omega }}}{{{\Omega _2}}}\sin ({\Omega _2}t + {\varphi _i}),\quad i = 2,3, \cdots ,n
\end{equation}
and
\begin{equation}
\begin{array}{l}
 {\omega _i}^2 = \frac{{{k_\omega }^2}}{2}\left( {\frac{{{A_i}^2}}{{{\Omega _1}^2}} + \frac{{{B_i}^2}}{{{\Omega _2}^2}}} \right) - \frac{{{k_\omega }^2{A_i}^2}}{{2{\Omega _1}^2}}\cos 2({\Omega _1}t + {\phi _i}) \\
 \quad \quad  - \frac{{{k_\omega }^2{B^2}}}{{2{\Omega _2}^2}}\cos 2({\Omega _2}t + {\varphi _i}),\quad i = 2,3, \cdots ,n \\
 \end{array}
\end{equation}
Substituting ${\omega _i}^2$ into the equation of the $ith$ oscillator, since the constant of ${\omega _i}^2$ dominates the natural frequency of the oscillator $x_i$, we get its natural frequency is
\begin{equation}
{\omega _{ni}} = \sqrt {\frac{{{k_\omega }^2}}{2}\left( {\frac{{{A_i}^2}}{{{\Omega _1}^2}} + \frac{{{B_i}^2}}{{{\Omega _2}^2}}} \right)} ,\quad i = 2,3, \cdots ,n
\end{equation}
Then, the approximated vibrational resonance condition for the $ith$ oscillator is
\begin{equation}
{\Omega _1} = \sqrt {\frac{{{k_\omega }^2}}{2}\left( {\frac{{{A_i}^2}}{{{\Omega _1}^2}} + \frac{{{B_i}^2}}{{{\Omega _2}^2}}} \right)} ,\quad i = 2,3, \cdots ,n
\end{equation}
From Eq.~(11) and Eq.~(16), we find that the vibrational resonance condition for the $ith$ oscillator and for the first oscillator is unified. As a result, they can be expressed by a general expression. Especially, when we know $\gamma = \Omega _2 / \Omega _1$, the vibrational resonance condition for each oscillator is
\begin{equation}
{\Omega _1} = \sqrt[4]{{\frac{{{k_\omega }^2}}{2}\left( {{A_i}^2 + \frac{{{B_i}^2}}{{{\gamma ^2}}}} \right)}},\quad i = 1,2, \cdots ,n
\end{equation}
Corresponding to the resonance peak, we also get the function of the learning rate $k_\omega$ and other important control parameters, i.e.,
\begin{equation}
{k_\omega } = {\Omega _1}^2\sqrt {\frac{2}{{{A_i}^2 + \frac{{{B_i}^2}}{{{\gamma ^2}}}}}} ,\quad i = 1,2, \cdots ,n
\end{equation}
where $A_1=A$ and $B_1=B$. Similarly, we get the relationship between $A_i$ and other parameters following the formula
\begin{equation}
{A_i} = \sqrt {\frac{{2{\Omega _1}^4}}{{{k_\omega }^2}} - \frac{{{B_i}^2}}{{{\gamma ^2}}}} ,\quad i = 1,2, \cdots ,n.
\end{equation}
Herein, when $i>1$, $A_i$ and $B_i$ should be extracted from the time series $f(t)+\epsilon x_{i-1}$. It is easy to numerically calculate $A_i$ and $B_i$ by the following formulas
\begin{equation}
{A_i} = \sqrt {{A_{is}}^2 + {A_{ic}}^2} ,\quad i = 2,3, \cdots ,n
\end{equation}
where
\begin{equation}
\left\{ \begin{array}{l}
 {A_{is}} = \frac{2}{{mT}}\int\limits_0^{mT} {[f} (t) + \epsilon {x_{i - 1}}]\sin ({\Omega _1}t){\rm{d}}t \\
 {A_{ic}} = \frac{2}{{mT}}\int\limits_0^{mT} {[f} (t) + \epsilon {x_{i - 1}}]\cos ({\Omega _1}t){\rm{d}}t \\
 \end{array} \right.,\quad i = 2,3, \cdots ,n
\end{equation}
and
\begin{equation}
{B_i} = \sqrt {{B_{is}}^2 + {B_{ic}}^2} ,\quad i = 2,3, \cdots ,n
\end{equation}
where
\begin{equation}
\left\{ \begin{array}{l}
 {B_{is}} = \frac{2}{{mT}}\int\limits_0^{mT} {[f} (t) + \epsilon {x_{i - 1}}]\sin ({\Omega _2}t){\rm{d}}t \\
 {B_{ic}} = \frac{2}{{mT}}\int\limits_0^{mT} {[f} (t) + \epsilon {x_{i - 1}}]\cos ({\Omega _2}t){\rm{d}}t \\
 \end{array} \right.,\quad i = 2,3, \cdots ,n
\end{equation}
\\
\indent Figure~3 shows that all four oscillators exhibit vibrational resonance. However, when the characteristic frequency $\Omega_{1}$ is large, resonance at $\Omega_{1}$ is difficult to achieve in a conventional Duffing oscillator or in a conventional coupled array. Introducing the learning rules markedly extends the frequency range over which resonance can occur, as evident in Fig.~3. For fixed $\Omega_{1}$, varying $B$ induces vibrational resonance-consistent with the classic VR mechanism. For fixed $B$, Fig.~3 also shows a resonance peak at a specific $\Omega_{1}$, with a profile resembling a traditional frequency-response curve. In addition, the cascading architecture enhances the signal, as reflected by the improved response across the oscillators number. White regions at each level denote parameter sets for which the response of the system diverges.\\
\begin{figure}%[h]
\includegraphics[width=0.5\textwidth]{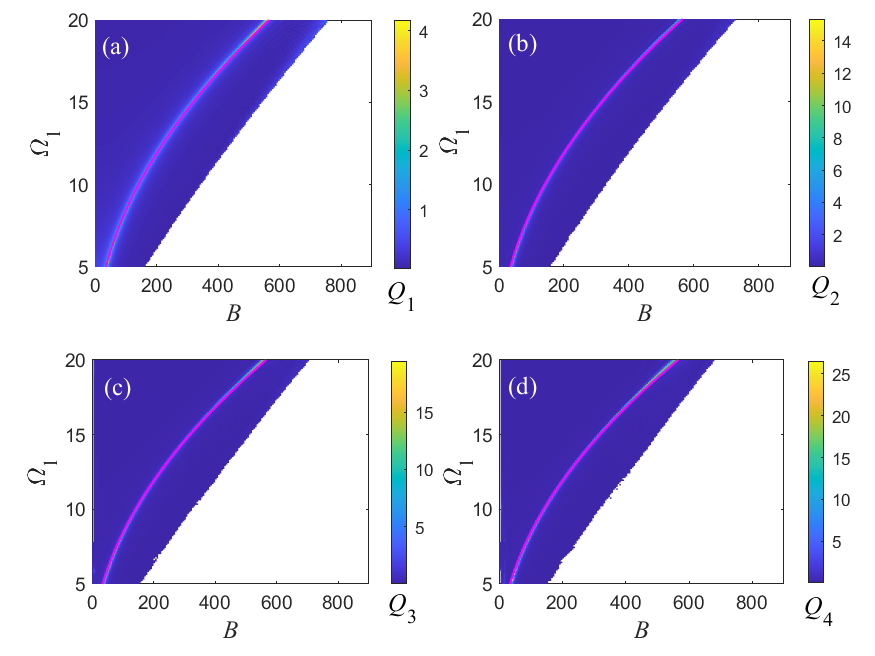}% Here is how to import EPS art
\caption{Influence of the characteristic frequency $\Omega_1$ on vibrational resonance of Eq.~(3) in the $B-\Omega_1$ plane. The continuous lines in purple color are the theoretical results and the numerical results are covered due to the consistent of the two kinds of results. The simulation parameters are $\zeta=0.25$, $b=1$, $\beta=10$, $k_\omega=10$, $A=0.1$, $\Omega_2=10\Omega_1$ and $\epsilon=1$. Panels (a)-(d), $Q_i$ versus $B$ and $\Omega_1$, $i=1$, 2, 3, 4 in turn. This figure corresponds to Fig.~A1 of the Appendix which removes the theoretical results of the resonance ridge line.}
\end{figure}
\indent Figure~4 illustrates the influence of the learning rate on vibrational resonance. As the learning rate $k_\omega$ increases, the value of $B$ required to induce vibrational resonance decreases. At the same time, larger $k_\omega$ values make the system more susceptible to divergence. The figure also confirms that vibrational resonance strengthens progressively as the number of coupled oscillators increases.
\begin{figure}%[h]
\includegraphics[width=0.5\textwidth]{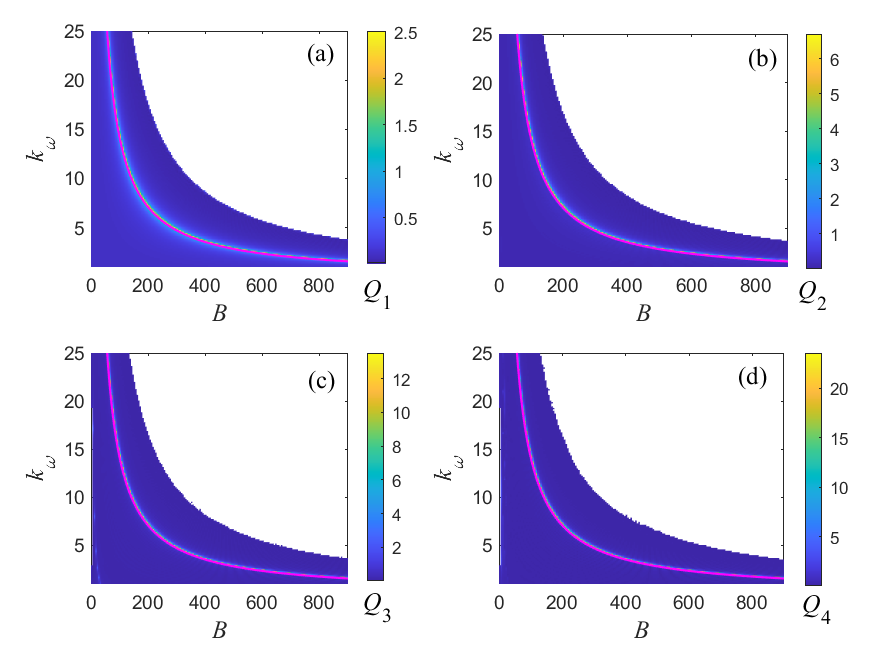}% Here is how to import EPS art
\caption{Influence of the learning rate $k_\omega$ on vibrational resonance of Eq.~(3) in the $B-k_\omega$ plane. The continuous lines in purple color are the theoretical results and the numerical results are covered due to the consistent of the two kinds of results. The simulation parameters are $\zeta=0.25$, $b=1$, $\beta=10$, $A=0.1$, $\Omega_1=10$, $\Omega_2=10\Omega_1$ and $\epsilon=1$. Panels (a)-(d) show $Q_i$ as a function of $B$ and $k_\omega$ for $i = 1, 2, 3, 4$, respectively. This figure corresponds to Fig.~A2 of the Appendix which removes the theoretical results of the resonance ridge line.}
\end{figure}
\\
\indent In addition, in Fig.~3 and Fig.~4, the approximated theoretical predictions for the occurrence of vibrational resonance are completely consistent with the numerical calculation results, indicating the validity of the approximated theoretical analysis. Further, to reveal the resonance structure clearly, corresponding to Fig.3 and Fig.4, we also give the plots that the curves of the theoretical results are removed and only the numerical results are retained in Fig.A1 and Fig.A2 of the Appendix.\\
\indent To examine in detail the relationship between the response amplitude $Q_i$ and the signal amplitude $B$ acting on each oscillator, we present Fig.~5. As shown in the figure, although the resonance peak differs for each oscillator, the value of $B$ at which the peak occurs decreases slightly as the number of oscillators increases.
\begin{figure}%[h]
\includegraphics[width=0.45\textwidth]{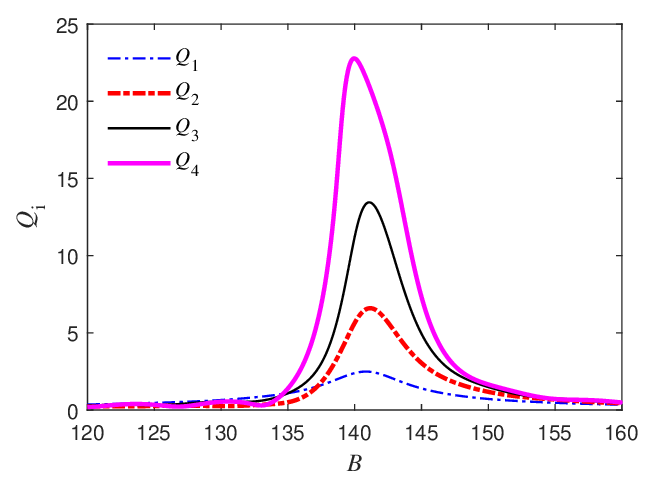}% Here is how to import EPS art
\caption{$Q$ versus $B$ of the first four oscillators of Eq.~(3). The simulation parameters are $\zeta=0.25$, $b=1$, $\beta=10$, $k_\omega=10$, $A=0.1$, $\Omega_1=10$, $\Omega_2=10\Omega_1$ and $\epsilon=1$.}
\end{figure}
\\
\indent Corresponding to the resonance region in Fig.~5, a set of time-series outputs is shown in Fig.~6. The results demonstrate that, from the first to the fourth oscillator, the characteristic frequency $\Omega_{1}$ becomes increasingly pronounced, indicating strong resonance. At the same time, suppression of other frequency components in the output is enhanced as the number of oscillators increases.
\begin{figure}
\includegraphics[width=0.5\textwidth]{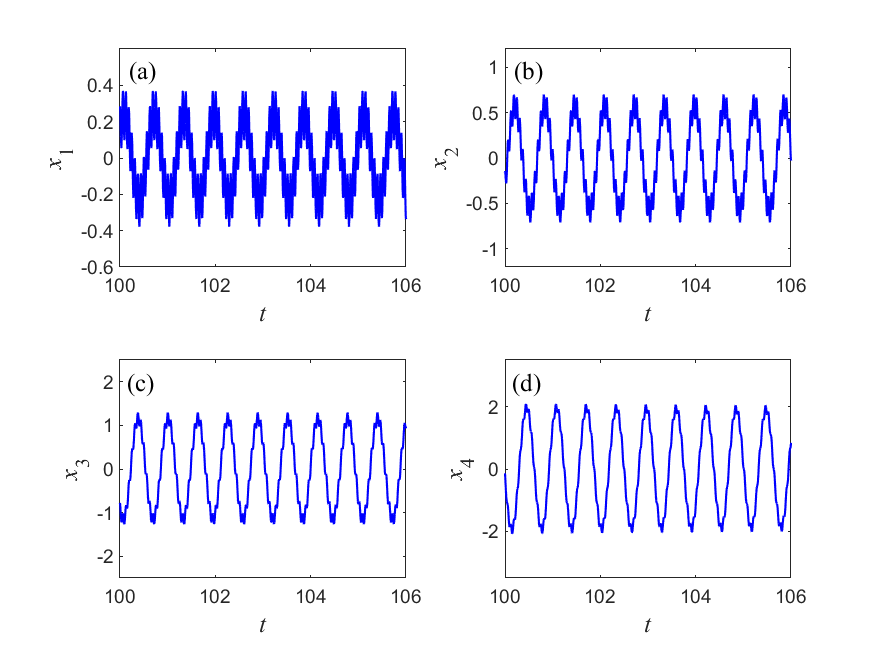}% Here is how to import EPS art
\caption{Time series of Eq.~(3) over a short interval during vibrational resonance. Panels (a)-(d) show the resonance responses of the first through fourth oscillators, respectively.  The simulation parameters are $\zeta = 0.25$, $b = 1$, $\beta = 10$, $A = 0.1$, $B = 140$, $\Omega_{1} = 10$, $\Omega_{2} = 10 \Omega_{1}$ and $\epsilon=1$.}
\end{figure}
\\
\indent In the above results, the signal is already strongly enhanced by an array of four coupled oscillators, so further discussion of increasing the number of oscillators is unnecessary.
\subsection{Influence of the signal amplitude $A$}
The influence of the signal amplitude $A$ on the degree of vibrational resonance for different oscillators are illustrated in Fig.~7. Although vibrational resonance focuses on the characteristic signal in weak case, in this figure, we extend the signal amplitude to a relative large scope. Some interesting phenomenons are shown in this figure. At first, vibrational resonance occurs in each oscillator, and the degree of the strongest resonance of each oscillator increases with the oscillators number. Second, for the first two oscillators, the maximal response amplitude does not show a significant change with the increase of the signal amplitude $A$. However, for the third and fourth oscillators, the maxial response amplitude decreases with the increase of amplitude $A$. In fact, we do not need to dwell on these issues, because vibrational resonance mainly targets weak characteristic signals. And for weak signals, the resonance effect of each level of oscillator satisfies the requirement. Further, the vibrational resonance region corresponding to the signal amplitudes $A$ and $B$ can also be predicated by Eq.~(19). In Fig.A3 of the Appendix, the consistent of the approximated theoretical resonance ridge lines with the numerical simulation results are verified effectively.
\begin{figure}[h]
\includegraphics[width=0.5\textwidth]{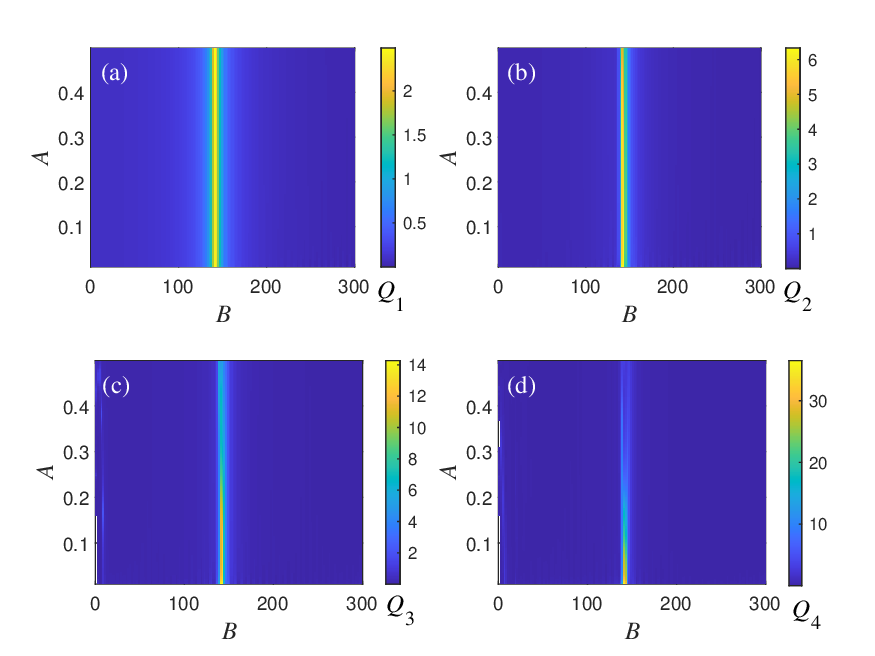}% Here is how to import EPS art
\caption{Influence of the signal amplitude $A$ on Vibrational resonance in the $B-A$ plane. The simulation parameters are $\zeta=0.25$, $b=1$, $\beta=10$, $k_\omega=10$, $\Omega_1=10$, $\Omega_2=10\Omega_1$ and $\epsilon=1$. Panels (a)-(d), $Q_i$ versus $B$ and $A$, $i=1$, 2, 3, 4 in turn. This figure corresponds to Fig.~A3 of the Appendix which adds the theoretical results of the resonance ridge line.}
\end{figure}
\subsection{Effect of the coupling strength $\epsilon$}
To show the effect of the coupling strength on vibrational resonance, we give Fig.7 which illustrates the plot of $Q$ versus the coupling strength $\epsilon$ and the signal amplitude $B$. For the first oscillator, it has nothing to do with the coupling strength, and this fact is indicated from both Eq.~(3) and Fig.~7(a). For the second oscillator, its vibrational resonance degree is much stronger than the first one, and the coupling strength enhance vibrational resonance. For the third and fourth oscillators, an appropriate coupling strength value makes the system achieve a satisfactory vibrational resonance. However, with the increase of $\epsilon$, the amplitude of the response diverges due to excessive size. It is the same as the divergence caused by excessive amplitude due to resonance in the traditional sense. As a result, we choose a the coupling strength within the appropriate range, e.g., $\epsilon=1$, in other simulation examples.
\begin{figure}[h]
\includegraphics[width=0.5\textwidth]{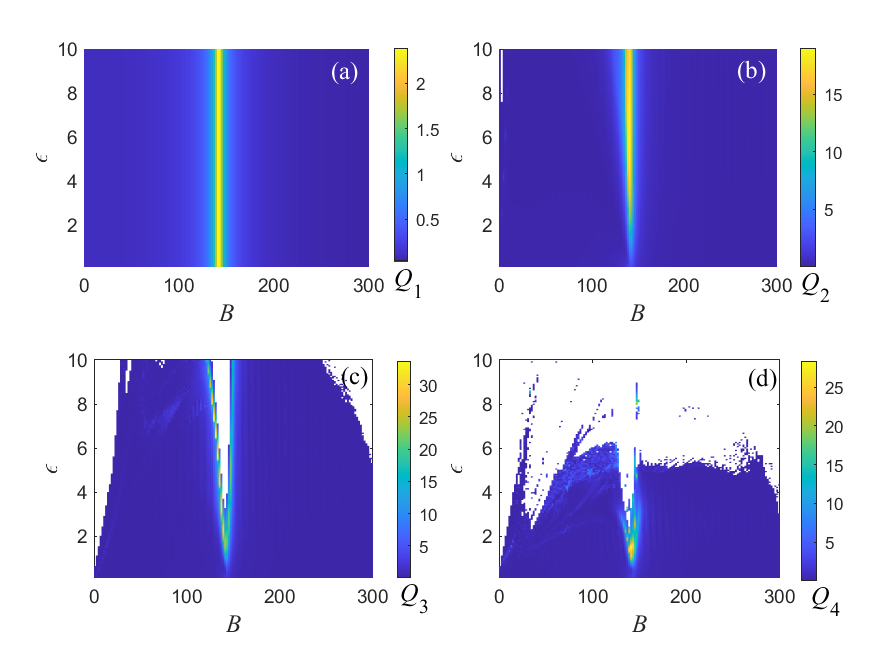}% Here is how to import EPS art
\caption{Influence of the coupling strength $\epsilon$ on vibrational resonance in the $B-\epsilon$ plane. The simulation parameters are $\zeta=0.25$, $b=1$, $\beta=10$, $k_\omega=10$, $A=0.1$, $\Omega_1=10$, $\Omega_2=10\Omega_1$. Panels (a)-(d), $Q_i$ versus $B$ and $\epsilon$, $i=1$, 2, 3, 4 in turn.}
\end{figure}
\subsection{Another kind of coupled self-learning Duffing oscillators}
\indent Herein, we further simplify Eq.(3) to get another type of coupled self-learning Duffing oscillators, which has a much simpler learning rule, i.e.,
\begin{equation}
\left\{ {\begin{array}{*{20}{c}}
   {{{\ddot x}_1} + 2\zeta {{\dot x}_1} + {\omega _1}^2{x_1} + b{x_1}^3 = \beta f(t)}  \\
   {{{\ddot x}_i} + 2\zeta {{\dot x}_i} + {\omega _i}^2{x_i} + b{x_i}^3 = \beta [f(t) + {x_{i - 1}}]}  \\
   {{{\dot \omega }_i} = {k_\omega }f(t),\quad i = 1,\:2, \cdots ,\:n}  \\
   {f(t) = A\cos ({\Omega _1}t) + B\cos ({\Omega _2}t)}  \\
\end{array}} \right.
\end{equation}
In this case, the learning rule for all oscillators are identical. Then, the vibrational resonance condition for all oscillators are given by the same formula
\begin{equation}
{\Omega _1} = \sqrt {\frac{{{k_\omega }^2}}{2}\left( {\frac{{{A^2}}}{{{\Omega _1}^2}} + \frac{{{B^2}}}{{{\Omega _2}^2}}} \right)}
\end{equation}
\\
\indent To make the paper concise and condensed, we only provide an example to illustrate that this simple learning rule is also effective for vibrational resonance, as shown in Fig.~9. This figure gives the theoretical and numerical results for vibrational resonance simultaneously, and the corresponding plots which removing the theoretical results are given in Fig.~A4 of the Appendix. The learning rule can also achieve a satisfactory result is due to the resonance occurs at the characteristic frequency $\Omega_1$. However, if we will investigate the nonlinear vibrational resonance, such as it occurs at the nonlinear frequencies which is a multiple or fractional times of $\Omega_1$ \cite{Ref49, Ref50, Ref51}, using the learning rules in Eq.~(24) is inappropriate. In other words, the learning rules in Eq.~(3) are in much general forms, and they can achieve the adaptation of the considered frequencies not only the fundamental characteristic frequency. The investigated frequencies must be contained in the learning rules. This work focuses on theoretical research, and the learning rule in Eq.~(3) makes the research more universal. The role played by the learning rule in Eq.~(24) can be considered as a special case of the learning rule in Eq.~(3). Further, in some cases, we do not know the excitation signals directly, but they are submerged by other noise and interference frequencies. Especially in coupled oscillator arrays, in order for each level of oscillator to function well, the excitation and learning rules should include the response of the previous oscillator. Considering these points, we still adopt the rules in Eq.~(3) in the following study. It is more reasonable and easier for future research on vibrational resonance at more complex frequency components, even for the unknown signals.
\begin{figure}[h]
\includegraphics[width=0.5\textwidth]{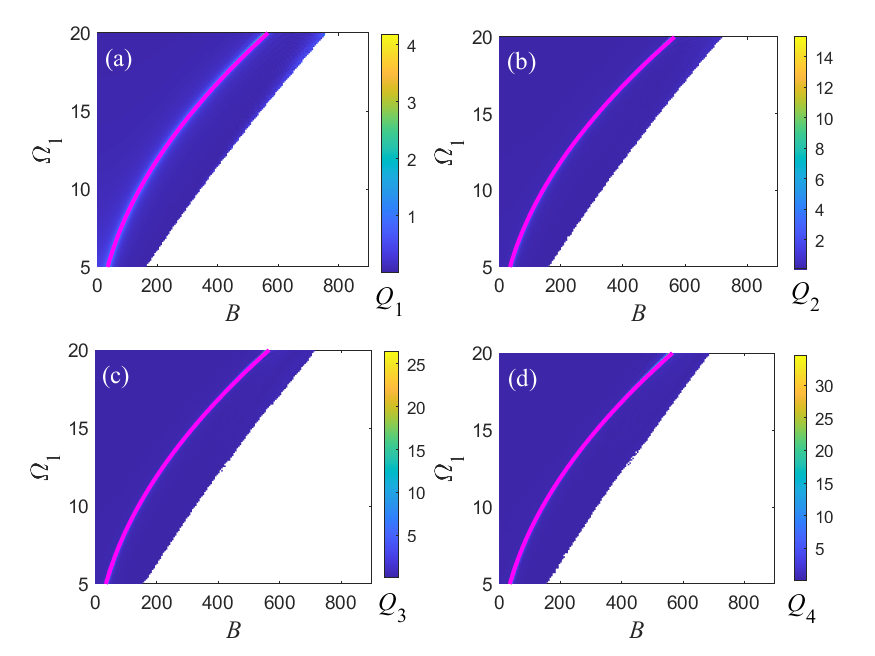}% Here is how to import EPS art
\caption{Influence of the characteristic frequency $\Omega_1$ on vibrational resonance of Eq.(24) in the $B-\Omega_1$ plane. The continuous lines in purple color are the theoretical results and the numerical results are covered due to the consistent of the two kinds results. The simulation parameters are $\zeta=0.25$, $b=1$, $\beta=10$, $A=0.1$, $k_\omega=10$, $\Omega_2=10\Omega_1$ and $\epsilon=1$. Panels (a)-(d), $Q_i$ versus $B$ and $\Omega_1$, $i=1$, 2, 3, 4 in turn. This figure corresponds to Fig.~A4 of the Appendix which removes the theoretical results of the resonance ridge line.}
\end{figure}
\section{Vibrational resonance in noisy environment}
\subsection{Simulated noisy signal denoising}
When noise is taken into account, Eq.~(3) in the case $\epsilon=1$ is modified as follows
\begin{equation}
\left\{ \begin{array}{l}
 {{\ddot x}_1} + 2\zeta {{\dot x}_1} + {\omega _1}^2{x_1} + b{x_1}^3 = \beta [f(t) + \xi (t)] \\
 {{\dot \omega }_1} = {k_\omega }[f(t) + \xi (t)] \\
 {{\ddot x}_i} + 2\zeta {{\dot x}_i} + {\omega _i}^2{x_i} + b{x_i}^3 = \beta [f(t) + \xi (t) + {x_{i - 1}}]\\
 {{\dot \omega }_i} = {k_\omega }\left[ {f(t) + \xi (t) + {x_{i - 1}}} \right],\quad i = 2,\;3, \cdots ,\;n \\
 f(t) = A\cos ({\Omega _1}t) + B\cos ({\Omega _2}t). \\
 \end{array} \right.
\end{equation}
The Gaussian white noise is directly added to the characteristic signal using the AWGN function of the MATLAB software, and the signal-to-noise ratio (SNR) of the noisy characteristic signal after adding noise is $SNR$. This operation is directly completed by the AWGN function of MATLAB. In practical applications, the collected signals are usually noisy, and the characteristic signal does not exist separately. Therefore, in the learning rule of Eq.~(26), the influence of noise is also considered.\\
\indent In Fig.~10, different values of $SNR$ are used in the input signal to verify the noise-reduction effect of the vibrational resonance algorithm, and the values of $Q_i$ are obtained by averaging over 50 trials. Compared with the noise-free case shown in Fig.~5, $Q_i$ decreases as the noise strength increases. Nevertheless, even under strong noise conditions, as illustrated in Figs.~10(a) and 10(b), the response still exhibits an obvious vibrational resonance phenomenon.
\begin{figure}
\includegraphics[width=0.5\textwidth]{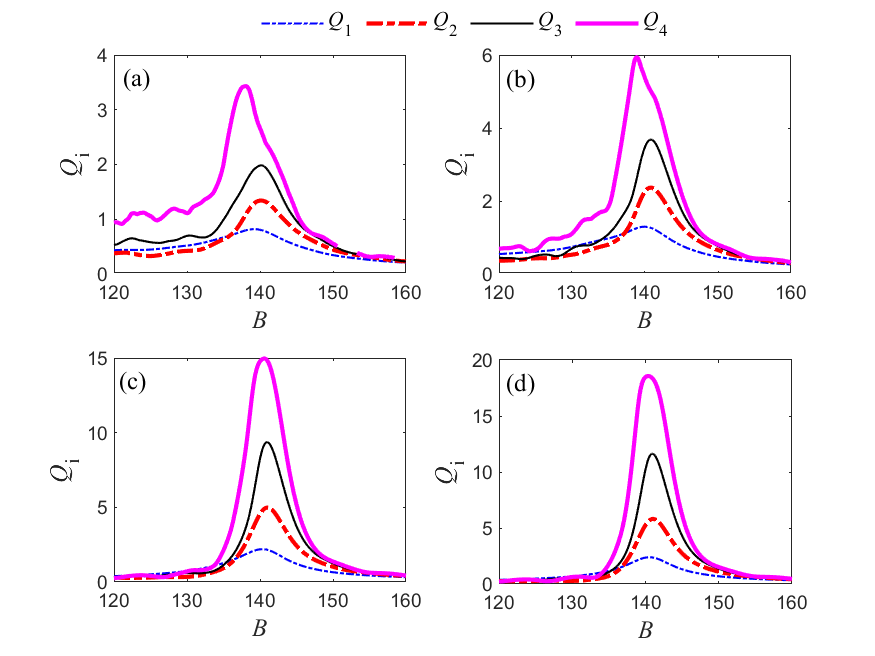}% Here is how to import EPS art
\caption{ Vibrational resonance in the noisy environment. The $SNR$ of the original noisy signal is different, (a) $SNR=-10$, (b) $SNR=-5$, (c) $SNR=5$, (d) $SNR=10$. Other simulation parameters are $\zeta=0.25$, $b=1$, $\beta=10$, $A=0.1$, $\Omega_1=10$ and $\Omega_2=10\Omega_1$.}
\end{figure}
\\
\indent Figure~11 illustrates the noise-reduction effect achieved by vibrational resonance in the self-learning coupled oscillators. In Fig.~11(a), the signal is completely submerged in noise, and no characteristic information can be discerned in the time series. When vibrational resonance is induced in the system-corresponding to the resonance peak in Fig.~10(c), time series of Figs.~11(b)-11(d) show the resonance response of the first, second, and fourth oscillators, respectively. The characteristic frequency components are clearly visible in the outputs and become more pronounced as the number of oscillators increases. Thus, Fig.~11 further demonstrates the feasibility of extracting and enhancing a weak characteristic signal from a strong noise background using the vibrational resonance algorithm. Compared with previous studies, the characteristic signal processed here lies at a higher frequency, confirming that the learning rules enhance the system's signal-processing
 capability, which is an advantage of frequency-adaptive coupled oscillators.
\begin{figure}[h]
\includegraphics[width=0.5\textwidth]{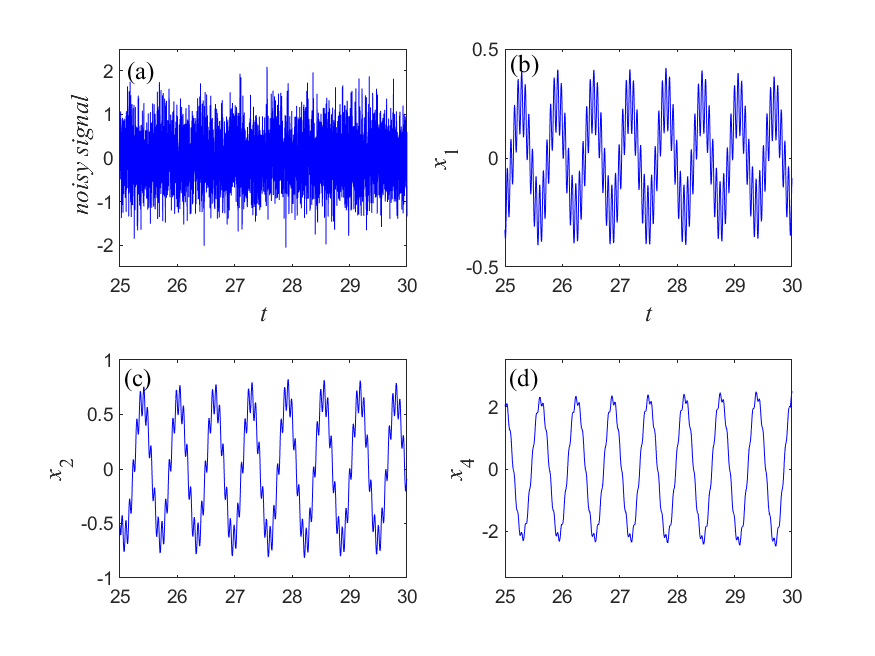}% Here is how to import EPS art
\caption{Comparison of the original noisy signal and the processed signals. (a) Original noisy signal. Panels (b)-(d), output time series of the first, second and fourth oscillator, respectively. The simulation parameters are  $\zeta=0.25$, $b=1$, $\beta=10$, $k_\omega=10$, $A=0.1$, $\Omega_1=10$, $\Omega_2=10\Omega_1$, $B=140.6$ and $SNR=5$.}
\end{figure}
\subsection{Experimental radio frequency signal denoising}
In this section, we take the electromagnetic wave signal of the passive radio frequency identification (RFID) system as an application. Sensing vibration information through a passive radio frequency tag has many advantages. For example, due to passive, wireless and lightweight properties, the tag is easy to install and can enable detection of previously unmeasured points due to the traditional sensors cannot be installed or line of sight limitations \cite{Ref52, Ref53, Ref54}. \\
\indent As shown in Fig.~12(a), the antenna emits electromagnetic waves to the passive tag, generating electromagnetic induction and activating the chip inside the tag. The built-in chip emits wireless electromagnetic waves. The tag is attached to the vibrating body and generates vibration together. The electromagnetic waves emitted by the tag carry vibration information, which is further received by the antenna and then settled and analyzed through a reader and computer. In free space, communication between reader and tag is based on the principle of electromagnetic wave propagation. Under ideal operating conditions, there is only a direct transmission path between the tag and the reader, and the signal does not undergo reflection or scattering, thus maintaining a high purity. However, in real situations, during the propagation of the radio frequency signal, metal structures or obstacles in the environment can cause multiple reflections, diffractions, and scattering, thus forming multiple propagation paths, as shown in Fig.~12(b). Due to differences in propagation distance and reflection frequency among different paths, it ultimately manifests as a multi-component coherent superposition at the reader, resulting in typical multipath effects. The multipath effect not only introduces strong noise interference into the radio frequency signal, but may also completely obscure the embedded characteristic information.\\
\indent We built a wireless RFID experimental system, as shown in Fig.~12(c). It mainly consists of a passive radio frequency tag, an antenna, a reader, and several auxiliary components. The tag is fixed on a board at the front end of the exciter. A signal generator produces a signal, which is amplified by a power amplifier before being fed into the exciter, driving it to vibrate according to a prescribed pattern. When the RFID system is activated, the antenna emits electromagnetic waves that power the passive tag through electromagnetic induction. As the exciter vibrates, the passive tag emits modulated electromagnetic waves outward. These signals carry information about the vibration characteristics of the tag. The antenna receives the emitted signals and forwards them to the reader, which demodulates them. The processed data are then transferred to a laptop, from which the vibration characteristics of the tag are finally obtained.\\
\indent During the operation of the RFID system, it is often disturbed by the surrounding environment, causing interference to the propagation of electromagnetic waves, and resulting in a large amount of complex noise in the collected signal. We set up a typical interference environment in Fig.~12(c), i.e., a multipath detection environment by placing a metal frame in the radio frequency signal propagation path. Under this environment, the electromagnetic waves undergo multiple reflections and diffractions in space, accompanied by partial occlusion effects, resulting in strong multipath interference noise. By adjusting the quantity and size of surfaces on the metal frame, the intensity of multipath interference can be effectively controlled, thereby changing the noise level. In the experimental process, the metal frame results in a noise environment. When all 6 metal rods are installed on the frame, more severe interference leads to stronger noise than the case of removing all 6 metal rods. In the experimental, we set the vibration frequency of the exciter is 3 Hz, i.e., 18.84 rad/s. In the following analysis, the vibrational resonance algorithm based on the self-learning Duffing oscillators described above is applied to reduce noise in the signal collected from the passive radio frequency tag.
\begin{figure*}%[h]
\center
\includegraphics[width=0.9\textwidth]{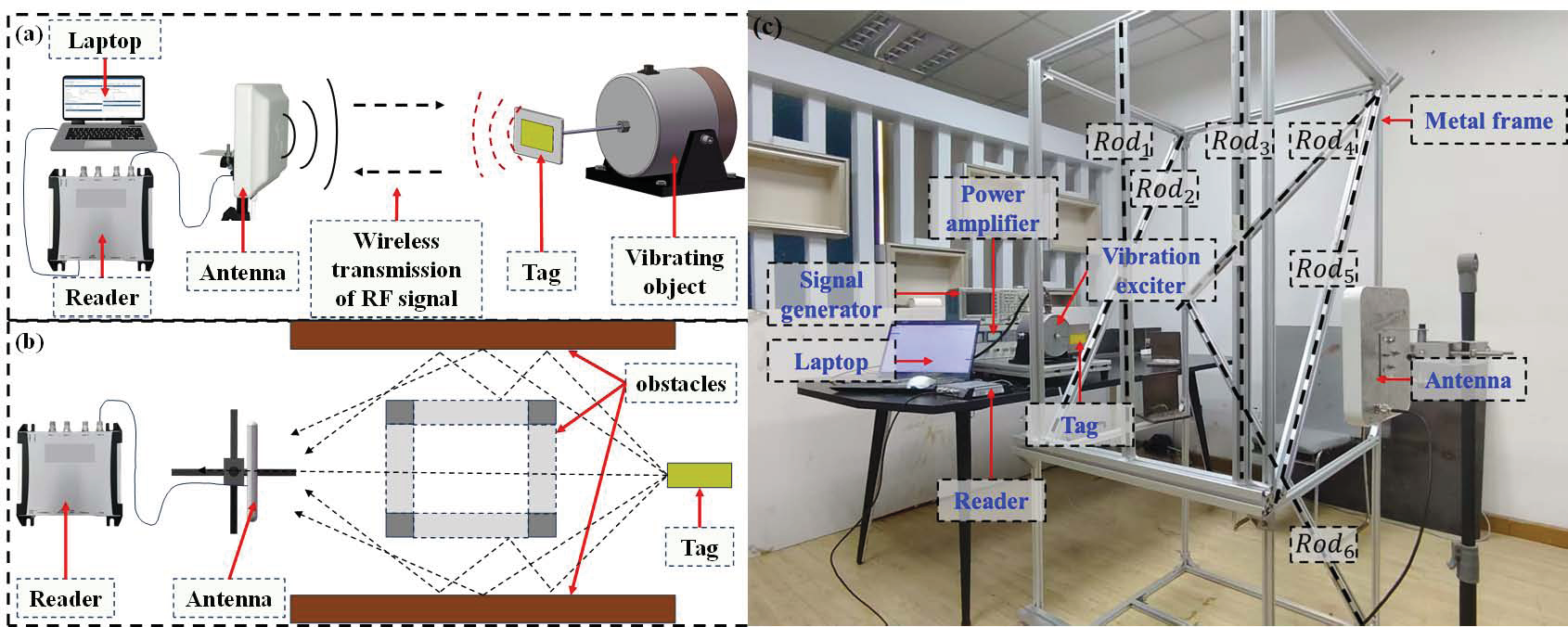}% Here is how to import EPS art
\caption{The principle of a RFID system and device of using passive radio frequency tag for vibration detection. (a) The basic principle of using passive radio frequency tag for vibration detection. (b) The principle of multipath effect causing noise interference to the radio frequency signal. (c) Experimental setup for vibration signal acquisition using a RFID system.}
\end{figure*}
\\
\indent Figure~13 clearly shows the effect of vibrational resonance on processing the signal collected in Fig.~12(c). The collected time series are directly processed without averaging $Q$ in the calculation, which leads to some fluctuations in the $Q_i$-$B$ curves. As seen in the figure, under different interference conditions, all signals collected from the passive radio-frequency tag contain strong background noise. Even after removing the metal rod groups from the frame, multipath interference continues to generate significant noise. For clarity in the analysis of the time series, we distinguish two cases: the frame without rods 1-6 is referred to as the weak-noise background case, while the frame with rods 1-6 is referred to as the strong-noise background case, according to the level of surrounding interference during signal acquisition.\\
\indent The effect of vibrational resonance on the radio frequency signal processing is illustrated in Fig.~13. From the output of the fourth oscillator, it is evident that the characteristic frequency in noisy environment is well recovered. At the same time, the amplitude of the wireless radio frequency signal varies during transmission due to environmental interference, leading to fluctuations in the oscillator output. Such fluctuations may also arise from the intrinsic dynamics of the oscillator array. Moreover, the entire dataset was processed completely, without omitting the transient response caused by the initial conditions, which explains the smaller amplitude observed at the beginning of the time series. Importantly, these amplitude fluctuations do not affect the practical applicability of the vibrational resonance method in engineering contexts. The results in Fig.~13 once again confirm that the vibrational resonance algorithm of the coupled self-learning Duffing oscillators is effective for recovering weak characteristic signals that buried in strong noise, achieving reliable processing performance.
\begin{figure}[h]
\center
\includegraphics[width=0.5\textwidth]{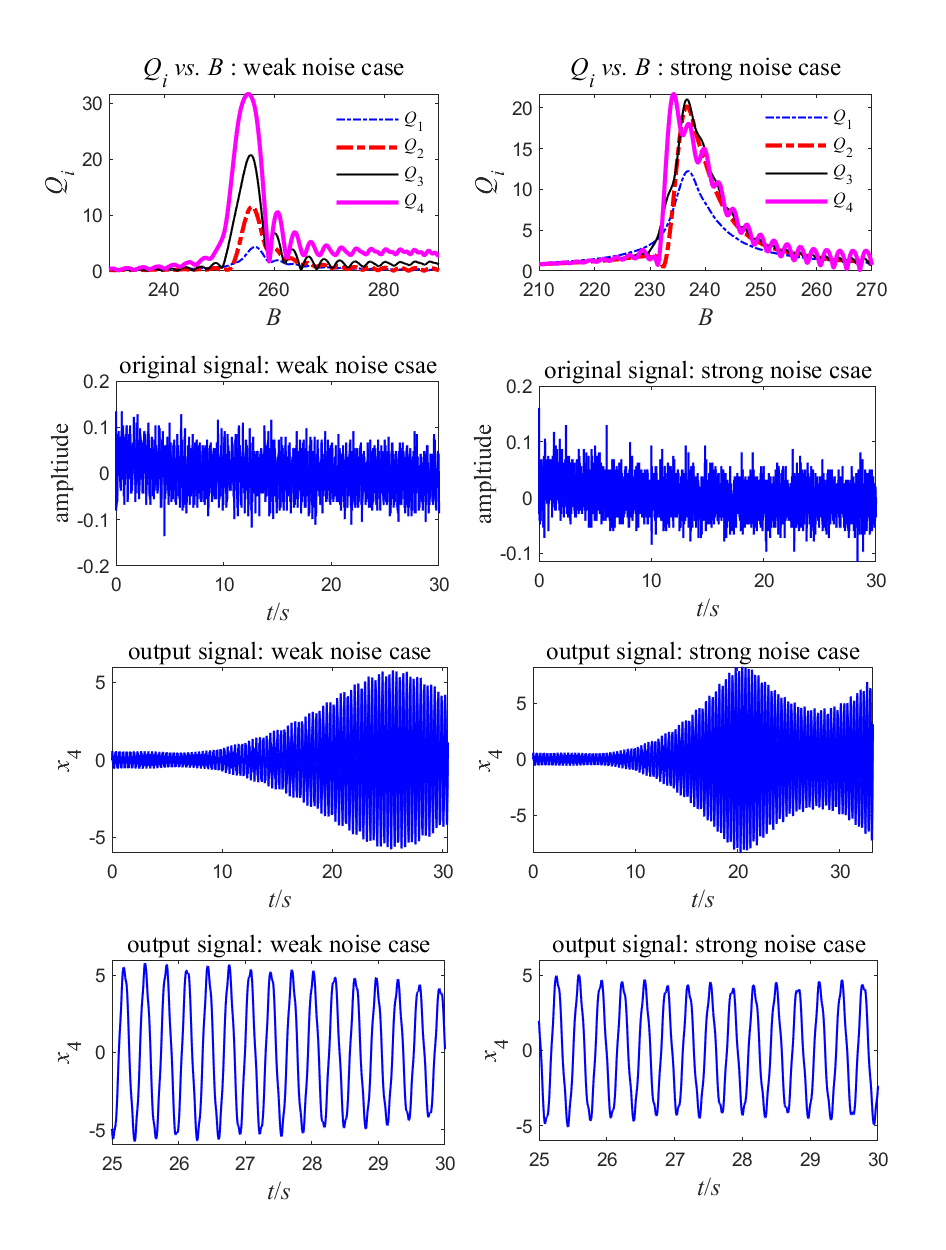}% Here is how to import EPS art
\caption{Experimental signal collected from the RFID system and the corresponding denoised result obtained by the coupled self-learning Duffing oscillators. The parameters for signal processing are $\zeta=0.45$, $b=1$, $\beta=10$, $k_\omega=10$ and $\Omega_2=90.43$. The output corresponds to the resonance peak of $Q_4$, and the optimal value $B$ is $B=255.4$ for the weak noise case and $B=235.4$ for the strong noise case.}
\end{figure}
\\
\indent In order to further verify the denoising effect and robustness advantages of vibrational resonance in frequency-adaptive learning Duffing system for feature frequency extraction under noise interference condition, herein, we compare the vibrational resonance method with two typical denoising algorithms, i.e., the wavelet transform and Kalman filter, as shown in Fig.~14. For wavelet denoising processing, Daubechies 4 (db4) is selected as the wavelet basis function, and the decomposition level is set to 5. In terms of algorithm parameter settings, the process noise covariance and measurement noise covariance of Kalman filtering are set to 0.001 and 1, respectively. The results indicate that the wavelet transform and Kalman filter can reduce noise to a certain extent when processing noisy signals, but their performance is highly sensitive to noise intensity. When the noise level is low, Kalman filtering can extract the signal waveform smoothly, while the effect of wavelet transform is relatively poor. However, as the noise intensity increases, the denoising ability of both two conventional methods significantly decreases, making it difficult to effectively extract feature signals. In contrast, the vibrational resonance method proposed in this work exhibits higher robustness and can achieve much more accurate feature extraction results under both weak and strong noise conditions. In fact, traditional denoising strategies such as wavelet transform and Kalman filter exhibit methodological limitations in some complex noise environments, and it is difficult to achieve reliable denoising when they are used singly \cite{Ref55, Ref56, Ref57}. Further, these traditional ways usually need to be combined with more advanced methods to enhance processing capabilities.
\begin{figure}%[h]
\center
\includegraphics[width=0.5\textwidth]{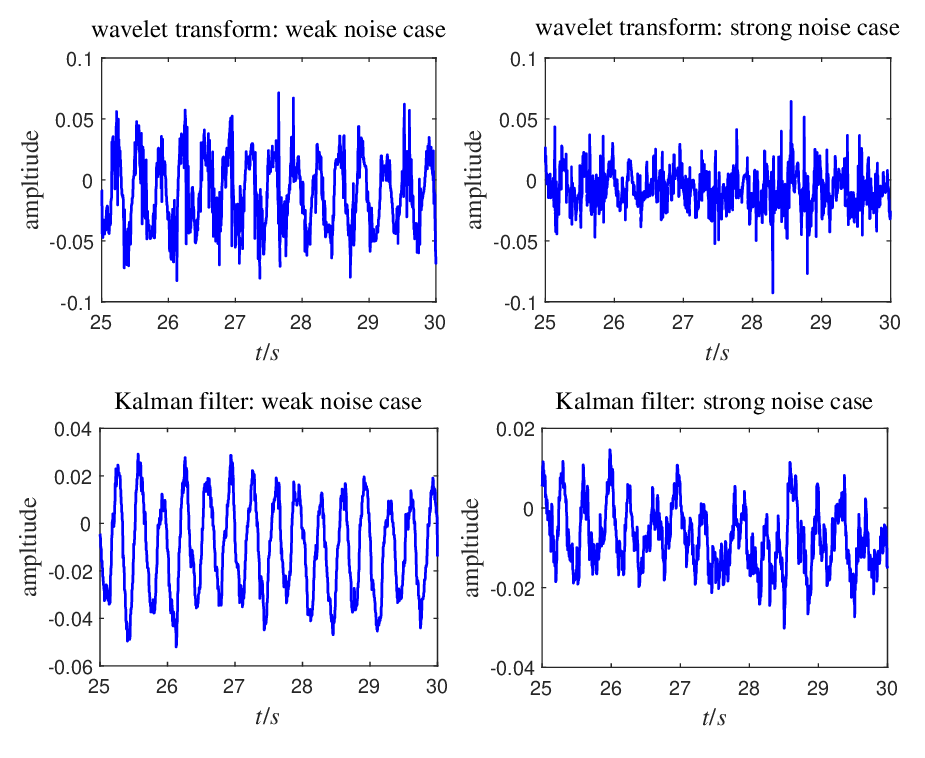}% Here is how to import EPS art
\caption{The signal denoising performance by wavelet transform and Kalman filter processing under weak and strong noise conditions. The processed original signals are the same with them in Fig.~13.}
\end{figure}
\subsection{Centrifugal pump vibration signal denoising}
\indent To further reveal the superiority and its application of the of vibrational resonance in frequency-adaptive learning Duffing oscillators in the noisy radio frequency processing, we give another validation scenario here. We built an experimental device for vibration detection of the centrifugal pump by the passive radio frequency tag, as shown in Fig.~15. Utilizing a RFID system to detect the vibration characteristic signal of the centrifugal pump not only verifies the effectiveness of the vibrational resonance method in the frequency-adaptive learning Duffing oscillators, but also covers the field of mechanical fault diagnosis and communication signal processing, reflecting the wide applicability of this new technology. The centrifugal pump of the experimental setup operates at a speed of 1500 revolutions per minute and with a corresponding rotating frequency of 25Hz. The experimental platform is equipped with a rotor misalignment fault, resulting in a characteristic frequency of 50Hz. The passive radio frequency tag is attached to the surface of the centrifugal pump near the coupling, and the vibration signal omits from the passive tag is received through an antenna to analyze the vibration characteristics of the centrifugal pump.
\begin{figure}%[h]
\center
\includegraphics[width=0.5\textwidth]{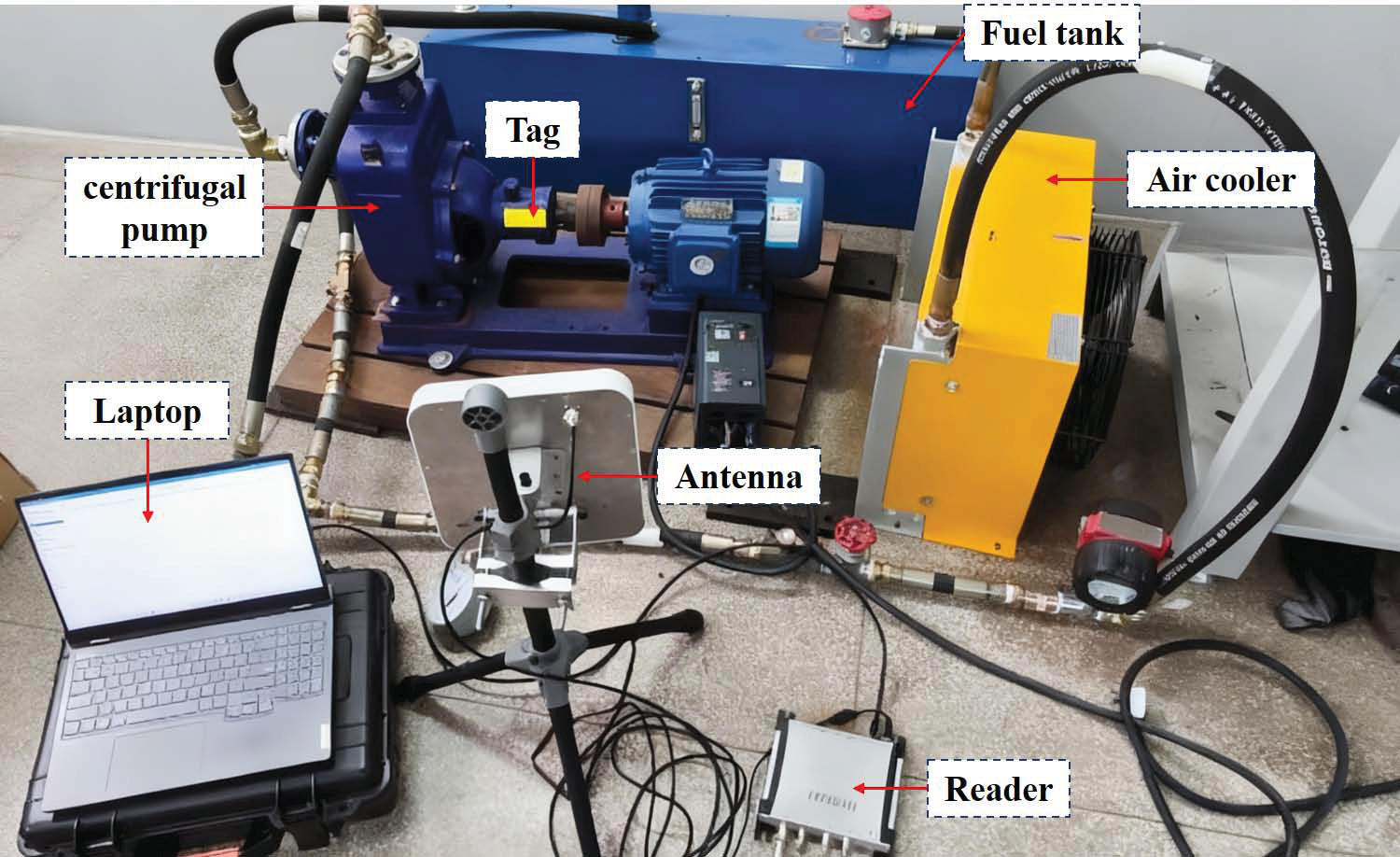}% Here is how to import EPS art
\caption{Fault detection experimental platform of a centrifugal pump based on passive radio frequency tag sensing.}
\end{figure}
\\
\indent Figure~16 shows the time domain and frequency domain comparison results between the original signal and it processed by vibrational resonance, wavelet transform, and Kalman filtering, respectively. Obviously, the original signal is affected by the background noise interference, resulting in a low SNR. The characteristic frequency (50Hz) induced by the rotor misalignment is submerged in the spectrum, making it difficult to achieve effective feature identification. After being processed by vibrational resonance of the coupled oscillators, the fault characteristic frequency is significantly improved in both the time waveform and frequency spectrum, and the signal exhibits higher purity and feature clarity. In contrast, although wavelet denoising and Kalman filtering can suppress some high-frequency noise, there are still many residual interference components in the processed signal, making it difficult to restore the periodic characteristics of the fault signal. The results show that the vibrational resonance method can accurately recover the second harmonic feature components, i.e., the misalignment fault characteristic frequency. Compared with wavelet transform and Kalman filtering, it exhibits stronger robustness and effectiveness in extracting a weak fault feature.
\begin{figure}[h]
\center
\includegraphics[width=0.5\textwidth]{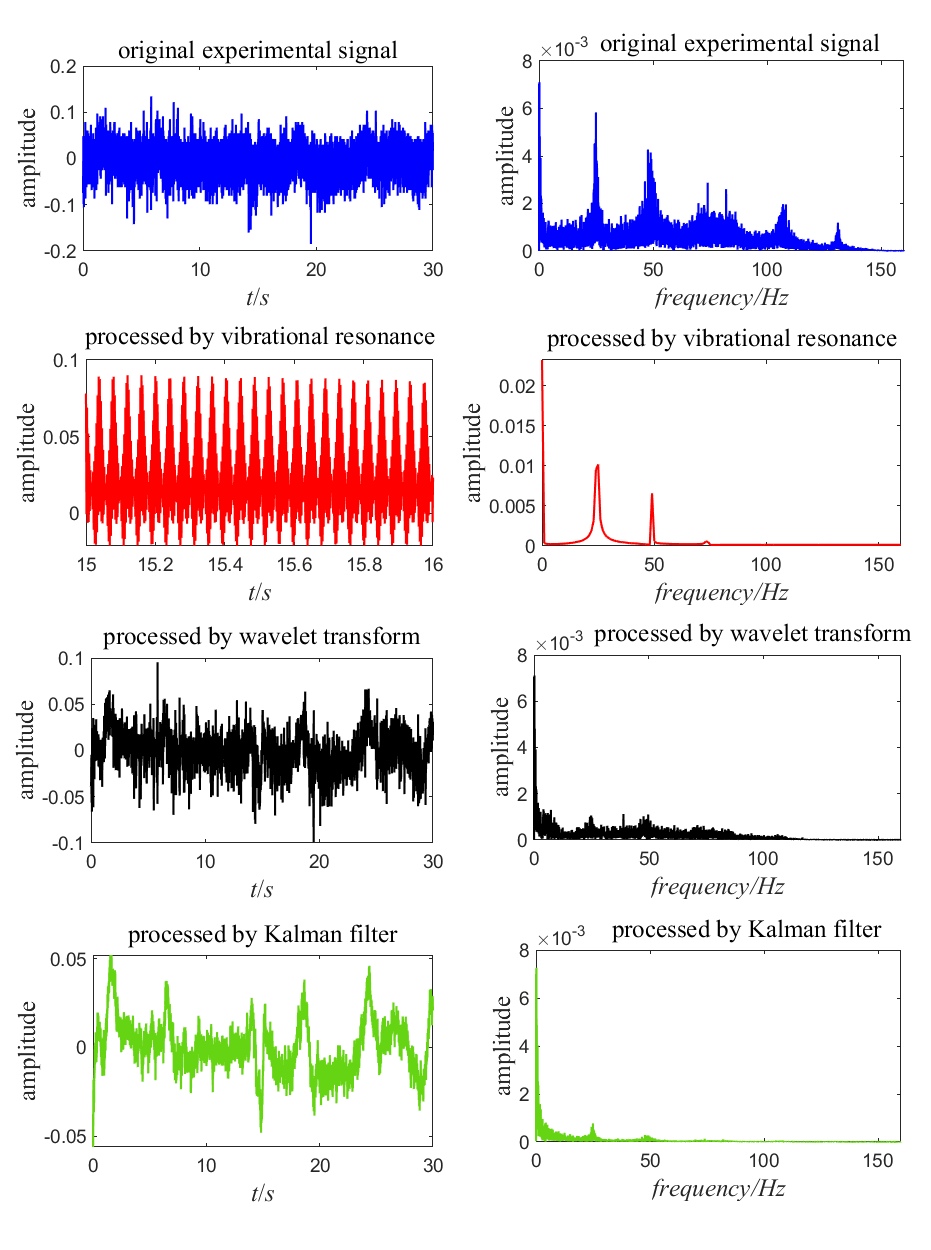}% Here is how to import EPS art
\caption{The noisy vibration signal collected from the platform of Fig.~15 and the results of misalignment fault extraction by vibrational resonance, wavelet transform and Kalman filter methods, respectively. Left panels: time-domain waveforms, right panels: frequency-domain spectrum.}
\end{figure}
\section{Conclusions}
This work proposes an array of coupled self-learning systems composed of frequency-adaptive Duffing oscillators. The approach addresses a key limitation of traditional vibrational resonance, which is typically effective only for low-frequency weak characteristic signals. By introducing parameter self-learning, vibrational resonance can be extended across a much broader range of characteristic frequencies. In addition, the theoretical condition for vibrational resonance is obtained and verified numerically under a group of different parameters. Furthermore, coupling multiple oscillators enhances the overall resonance effect compared to a single oscillator. By adjusting the learning rate and the amplitude of the high-frequency auxiliary signal, the resonance response at high characteristic frequencies can be effectively tuned. The vibrational resonance method not only amplifies pure characteristic signals but also successfully extracts weak signals buried in strong noise, demonstrating robust denoising and enhancement capabilities. Its effectiveness has been validated through both numerical simulations and experimental data obtained from two RFID systems operating under diverse interference conditions.\\
\indent Compared with previous research, the innovation of this work lies in the realization of vibrational resonance at high frequencies through simple learning rules, thereby avoiding the problem of parameter tuning required in earlier theoretical frameworks. At the same time, the proposed system and method show clear advantages in signal processing under strong noise backgrounds. This approach offers a new strategy for handling weak high-frequency characteristic signals in noisy environments, outperforming conventional vibrational resonance and other conventional methods such as wavelet transform and Kalman filter.\\
\indent In the future, three aspects of work should be valued. First, develop analytical methods to achieve more comprehensive analytical results of frequency-adaptive learning systems. Second, deeply study learning rules to enable them to adapt to more complex signals. Third, design corresponding hardware and implement further engineering applications of adaptive learning systems in the device. We believe that the main results of this work not only extend the existing achievements of vibrational resonance but will also generate potential engineering application values.
\section*{CRediT authorship contribution statement}
{\bf Jianhua Yang}: Conceptualization; Formal analysis; Methodology; Writing - original draft; Funding acquisition. {\bf Litai Lou}: Data curation; Methodology; Software. {\bf Shangyuan Li}: Data curation; Software. {\bf Zhongqiu Wang}: Data curation; Software. {\bf Miguel A. F. Sanju\'an}: Conceptualization; Funding acquisition; Writing - review \& editing.%
\section*{Declaration of competing interest}
The authors declare that they have no known competing financial interests or personal relationships that could have appeared to influence the work reported in this paper.
\section*{Acknowledgments}
The project was supported by the National Natural Science Foundation of China (Grant No. 12472036), the Priority Academic Program Development of Jiangsu Higher Education Institutions, and the Spanish State Research Agency (AEI) and the European Regional Development Fund (ERDF, EU) under Project No. PID2023-148160NB-I00.
\section*{Data Availability Statement}
The data that support the findings of this study are available from the corresponding author upon reasonable request.
\section*{Appendix}
%\section*{Appendix A}
\captionsetup[figure]{labelformat=simple, labelsep=colon}
\renewcommand{\thefigure}{A\arabic{figure}}
\setcounter{figure}{0}
To show the resonance structures obtained by the numerical calculations clearly and directly, we remove the resonance ridge lines that obtained by approximated theoretical predications from Eq.~(17), and get the results illustrated by Fig.~A1.
\begin{figure}[h]
%\center
\includegraphics[width=0.5\textwidth]{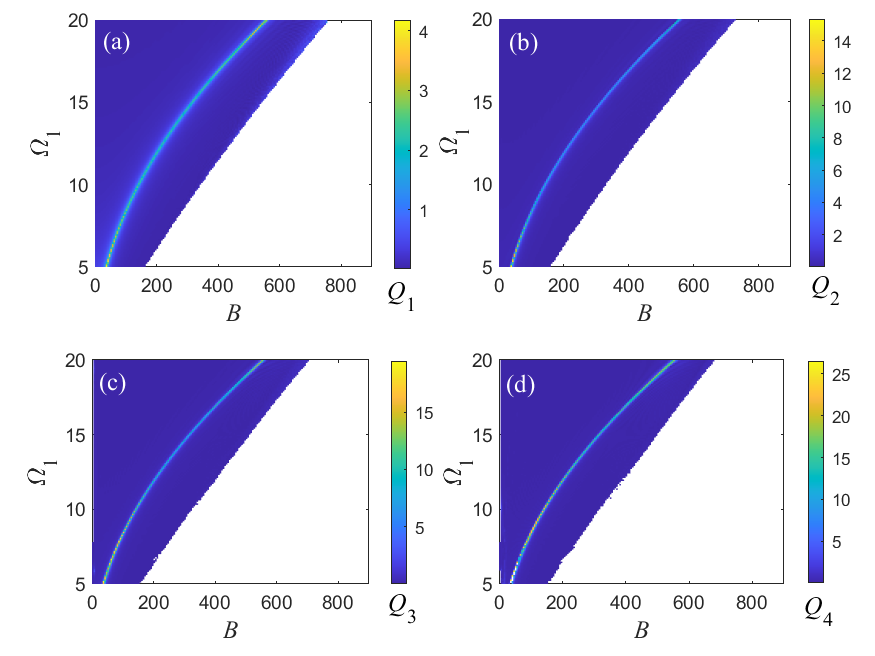}% Here is how to import EPS art
\\
{Figure A1: This figure corresponds to Fig.3 but the curves obtained by the approximated theoretical predications are removed.}
\end{figure}
\\
\indent Similarly, when we remove the resonance ridge lines of the approximated theoretical results of Eq.(18) in Fig.2, we get the numerical simulation results of the response amplitude in Fig.A2, which shows the resonance structures clearly.
\begin{figure}[h]
%\center
\includegraphics[width=0.5\textwidth]{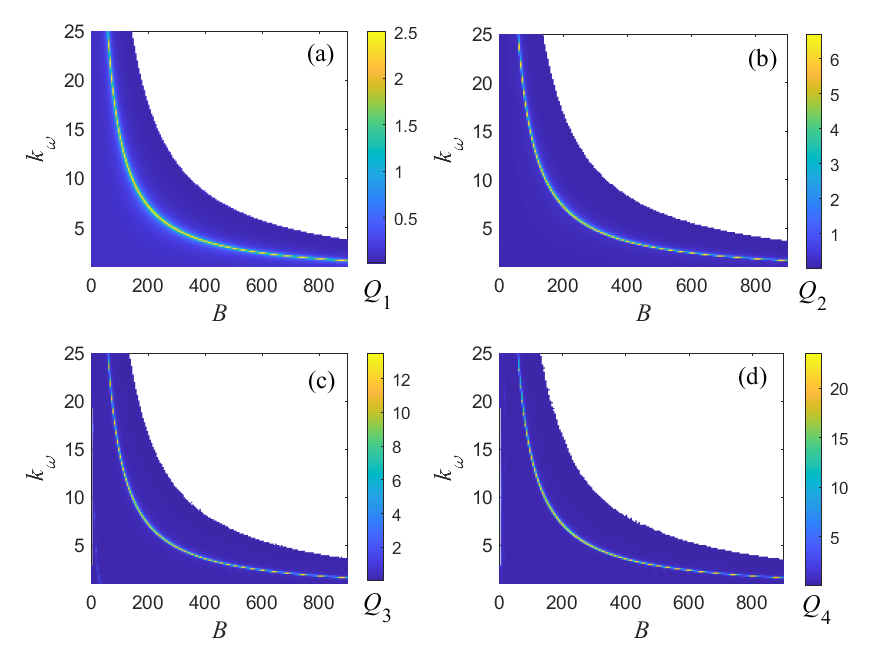}% Here is how to import EPS art
\\
{Figure A2: This figure corresponds to Fig.4 but the curves obtained by the approximated theoretical predications are removed.}
\end{figure}
\\
\indent Consistent of the theoretical resonance ridge line with the numerical simulation results is illustrated in Fig.A3, which corresponds to Fig.7 in the main text but both the theoretical results and the analytical results are plotted together.
\begin{figure}[h]
%\center
\includegraphics[width=0.5\textwidth]{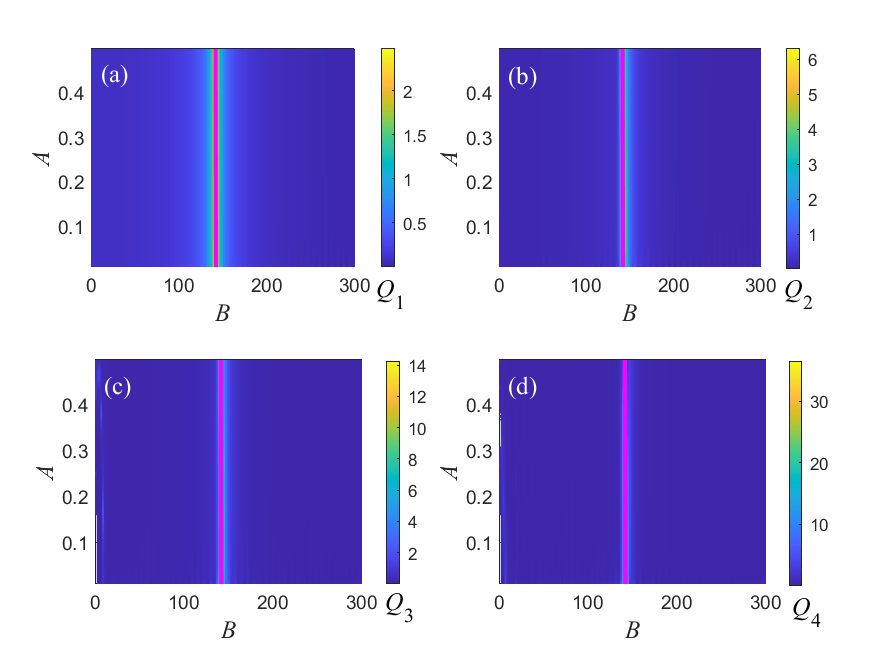}% Here is how to import EPS art
\\
{Figure A3: This figure corresponds to Fig.7 and the curves obtained by the approximated theoretical predications are also plotted in the figure. The continuous lines in purple color are the theoretical results and the numerical results are covered due to the consistent of the two kinds of results.}
\end{figure}
\\
\indent Corresponding to Fig.9, when the theoretical predication curves are removed, we obtain Fig.A4. Comparing Fig.A4 with Fig.A1, we find the response amplitude improves a little under the frequency-adaptive learning rule in Eq.(24).
\begin{figure}[h]
%\center
\includegraphics[width=0.5\textwidth]{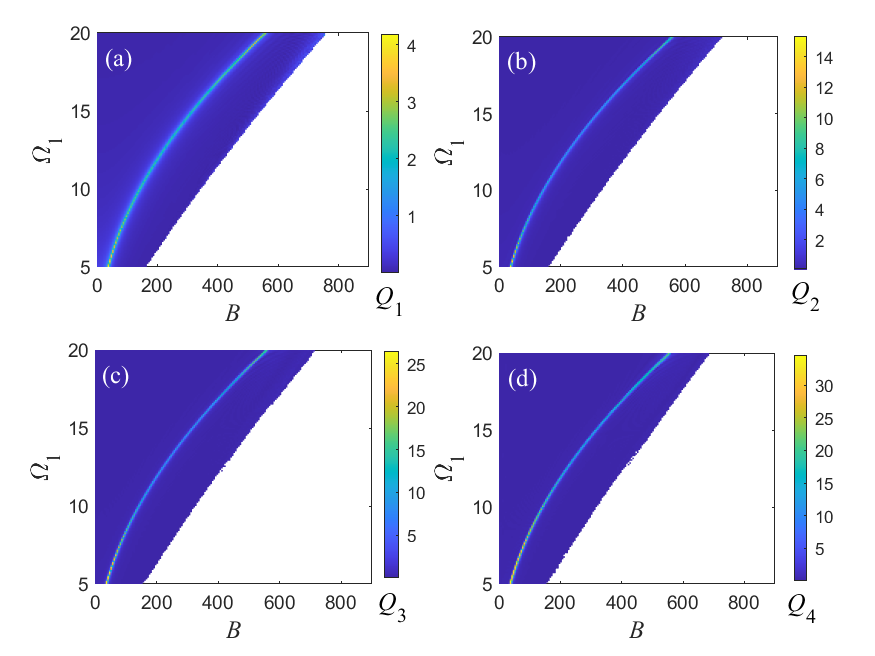}% Here is how to import EPS art
\\
{Figure A4: This figure corresponds to Fig.9 but the curves obtained by the approximated theoretical predications are removed.}
\end{figure}

\section*{References}
\begin{thebibliography}{00}

%% For numbered reference style
\bibitem{Ref1}
W. Moy, I. Ahmed, P.W. Chiu, J. Moy, S.S. Sapatnekar, C.H. Kim, A 1,968-node coupled ring oscillator circuit for combinatorial optimization problem solving, Nat. Electron. 5(5) (2022) 310-317.
\bibitem{Ref2}
M. Graber, K. Hofmann, An integrated coupled oscillator network to solve optimization problems, Commun. Eng. 3(1) (2024) 116.
\bibitem{Ref3}
G. Csaba, W. Porod, Coupled oscillators for computing: A review and perspective, Appl. Phys. Rev. 7(1) (2020) 011302.
\bibitem{Ref4}
V.V. Semenov, The impact of nonlocal coupling on deterministic and stochastic wavefront propagation in an ensemble of bistable oscillators, Phys. Lett. A 532 (2025) 130189.
\bibitem{Ref5}
A. Pikovsky, M. Rosenblum, Dynamics of globally coupled oscillators: Progress and perspectives, Chaos 25(9) (2015) 097616.
\bibitem{Ref6}
M.A. Budroni, G. Pagano, D. Conte, B. Paternoster, R. D'ambrosio, S. Ristori, A. Abou-Hassan, F. Rossi, Synchronization scenarios induced by delayed communication in arrays of diffusively coupled autonomous chemical oscillators, Phys. Chem. Chem. Phys. 23(32) (2021) 17606-17615.
\bibitem{Ref7}
C.H. Kim, J. Park, Y.J. Kim, S. Park, S. Boccaletti, B. Kahng, Cluster-mediated synchronization dynamics in globally coupled oscillators with inertia, Chaos Soliton Fract. 196 (2025) 116281.
\bibitem{Ref8}
T. Gong, J. Yang, S. Liu, H. Liu, Non-stationary feature extraction by the stochastic response of coupled oscillators and its application in bearing fault diagnosis under variable speed condition, Nonlinear Dynam. 108(4) (2022) 3839-3857.
\bibitem{Ref9}
S. Cheng, Y. Zheng, Y. Zhang, X. Liu, M. Yi, L. Lu, Weak signal amplification induced by oscillator diversity in a globally coupled bistable system, Chaos 35(9) (2025) 093136.
\bibitem{Ref10}
H. Liu, L. Wu, Y. Zheng, S. Cheng, X. Liu, L. Lu, Weak signal amplification induced by feedback coupling in a Y-shaped chain, Eur. Phys. J-Spec. Top. (2025) doi: 10.1140/epjs/s11734-025-01725-6.
\bibitem{Ref11}
Z. Yan, J.L. Guirao, T. Saeed, H. Chen, X. Liu, Analysis of stochastic resonance in coupled oscillator with fractional damping disturbed by polynomial dichotomous noise, Nonlinear Dynam. 110(2) (2022) 1233-1251.
\bibitem{Ref12}
S. Gao, N. Gao, B. Kan, H. Wang, Stochastic resonance in coupled star-networks with power-law heterogeneity, Physica A 580 (2021) 126155.
\bibitem{Ref13}
J. Zhu, M.E. Yamakou, Self-induced-stochastic-resonance breathing chimeras, Phys. Rev. E 108(2) (2023) L022204.
\bibitem{Ref14}
J. Song, X. Shi, J.H. Wu, T. Zheng, Z. Song, Nonlinear multi-order coupled stochastic resonance modeling under extremely low signal-to-noise ratios, Mech. Syst. Signal Pr. 224 (2025) 112208.
\bibitem{Ref15}
R. Zhang, L. Meng, L. Yu, S. Shi, H. Wang, Collective dynamics of fluctuating-damping coupled oscillators in network structures: Stability, synchronism, and resonant behaviors, Physica A 638 (2024) 129628.
\bibitem{Ref16}
K. Murali, S. Sinha, Coupling induced logical stochastic resonance, Phys. Lett. A 382(24) (2018) 1581-1585.
\bibitem{Ref17}
P.S. Landa, P.V.E. McClintock, Vibrational resonance, J. Phys. A: Math. Gen. 33(45) (2000) L433-L438.
\bibitem{Ref18}
V.M Gandhimathi, S. Rajasekar, J. Kurths, Vibrational and stochastic resonances in two coupled overdamped anharmonic oscillators, Phys. Lett. A 360(2) (2006) 279-286.
\bibitem{Ref19}
C. Yao, M. Zhan, Signal transmission by vibrational resonance in one-way coupled bistable systems, Phys. Rev. E 81(6) (2010) 061129.
\bibitem{Ref20}
E. Ullner, A. Zaikin, J. Garc\'ia-Ojalvo, R. Bascones, J. Kurths, Vibrational resonance and vibrational propagation in excitable systems, Phys. Lett. A 312(5-6) (2003) 348-354.
\bibitem{Ref21}
B. Deng, J. Wang, X. Wei, Effect of chemical synapse on vibrational resonance in coupled neurons, Chaos 19(1) (2009) 013117.
\bibitem{Ref22}
J. Li, X. Cheng, S. Zhang, Z. Meng, L. Cao, Fault feature extraction method of rolling bearings based on coupled resonance system with vibrational resonance-assisted enhanced stochastic resonance, Mech. Syst. Signal Pr. 208 (2024) 111069.
\bibitem{Ref23}
L. Xiao, J. Tang, X. Zhang, T. Xia, Weak fault detection in rotating machineries by using vibrational resonance and coupled varying-stable nonlinear systems, J. Sound Vib. 478 (2020) 115355.
\bibitem{Ref24}
A. Calim, T. Palabas, M. Uzuntarla, Stochastic and vibrational resonance in complex networks of neurons, Philos. T. R. Soc. A 379(2198) (2021) 20200236.
\bibitem{Ref25}
J. Yang and M.A.F. Sanju\'an, Aperiodic Resonances-Theory and Applications, Springer Nature, 2025.
\bibitem{Ref26}
J. Yang, S. Rajasekar, M.A.F. Sanju\'an, Vibrational resonance: A review, Phys. Rep. 1067 (2024) 1-62.
\bibitem{Ref27}
J. H. Yang, M. A. F. Sanju\'an, H. G. Liu, Enhancing the weak signal with arbitrary high-frequency by vibrational resonance in fractional-order Duffing oscillators, J. Comput. Nonlin. Dyn. 12(5) (2017) 051011.
\bibitem{Ref28}
T. Gong, J. Yang, M.A.F. Sanju\'an, H. Liu, Z. Shan, Vibrational resonance by using a real-time scale transformation method, Phys. Scripta 97(4) (2022) 045207.
\bibitem{Ref29}
J.R. Yang, C.J. Wu, J.H. Yang, H.G. Liu, On the weak signal amplification by twice sampling vibrational resonance method in fractional duffing oscillators, J. Comput. Nonlin. Dyn. 13(3) (2018) 031009.
\bibitem{Ref30}
E. Perkins, The Duffing adaptive oscillator, Nonlinear Dynam. 113(4) (2025) 2987-3000.
\bibitem{Ref31}
E. Perkins, Comparison of stochastic adaptation and stochastic resonance, Phys. Rev. E 110(6) (2024) 064225.
\bibitem{Ref32}
Y. Pan, F. Duan, F. Chapeau-Blondeau, L. Xu, D. Abbott, Study of vibrational resonance in nonlinear signal processing, Philos. T. R. Soc. A 379(2192) (2021) 20200235.
\bibitem{Ref33}
Y. Ren, Y. Pan, F. Duan, F. Chapeau-Blondeau, D. Abbott, Exploiting vibrational resonance in weak-signal detection, Phys. Rev. E 96(2) (2017) 022141.
\bibitem{Ref34}
X. Fu, S. Wu, C. Li, M. Xie, Z. Duan, B. Fan, Vibrational resonance and antiresonance in a membrane-in-the-middle optomechanical system, Chaos Soliton Frac 199 (2025) 116652.
\bibitem{Ref35}
J. Li, X. Cheng, S. Zhang, Z. Meng, L. Cao, Fault feature extraction method of rolling bearings based on coupled resonance system with vibrational resonance-assisted enhanced stochastic resonance, Mech. Syst. Signal Pr. 208 (2024) 111069.
\bibitem{Ref36}
J. Gao, J. Yang, D. Huang, H. Liu, S. Liu, Experimental application of vibrational resonance on bearing fault diagnosis, J. Braz. Soc. Mech. Sci. Eng. 41(1) (2019) 6.
\bibitem{Ref37}
L. Xiao, R. Bajric, J. Zhao, J. Tang,  X. Zhang, An adaptive vibrational resonance method based on cascaded varying stable-state nonlinear systems and its application in rotating machine fault detection, Nonlinear Dynamics, 103(1) (2021) 715-739.
\bibitem{Ref38}
S. Wang, B. Lu, Detecting the weak damped oscillation signal in the agricultural machinery working environment by vibrational resonance in the duffing system, J. Mech. Sci. Technol 36(12) (2022) 5925-5937.
\bibitem{Ref39}
S. Morfu, B.I. Usama, P. Marqui\'e, On some applications of vibrational resonance on noisy image perception: the role of the perturbation parameters, Philos. T. R. Soc. A 379(2198) (2021) 20200240.
\bibitem{Ref40}
S. Morfu, B.I. Usama, P. Marqui\'e, Perception enhancement of subthreshold noisy image with vibrational resonance, Electron. Lett. 55(11) (2019) 650-652.
\bibitem{Ref41}
M. Coccolo, G. Litak, J.M. Seoane, M.A.F. Sanju\'an, Energy harvesting enhancement by vibrational resonance, Int. J. Bifurcat. Chaos 24(6) (2014) 1430019.
\bibitem{Ref42}
I.A. Khovanov, The response of a bistable energy harvester to different excitations: the harvesting efficiency and links with stochastic and vibrational resonances, Philos. T. R. Soc. A  379(2198) (2021) 20200245.
\bibitem{Ref43}
T. Yang, Z. Huang, J. Xie, J. Liu, S. Li, B. Jiang, G. Zhu, X. Jing, Dynamic analysis and energy harvesting of double nonlinear stiffness vibration isolator, Eng. Struct. 332 (2025) 120028.
\bibitem{Ref44}
T. Yang, J. Xie, Z. Huang, J. Liu, H. Luo, X. Jing, Bio-inspired vibration isolator with triboelectric nanogenerator for self-powered monitoring, Mech. Syst. Signal Pr. 223 (2025) 111854.
\bibitem{Ref45}
T. Yang, J. Liu, H. Luo, Z. Li, Improving the performance of nonlinear isolator through triboelectric nanogenerator damper integrating energy harvesting, Energy 293 (2024) 130722.
\bibitem{Ref46}
Y. Cui, T. Yang, H. Luo, Z. Li, X. Jing, Jellyfish-inspired bistable piezoelectric-triboelectric hybrid generator for low-frequency vibration energy harvesting, Int. J. Mech. Sci. 279 (2024) 109523.
\bibitem{Ref47}
L. Righetti, J. Buchli, A.J. Ijspeert, Dynamic hebbian learning in adaptive frequency oscillators, Physica D 216(2) (2006) 269-281.
\bibitem{Ref48}
S. Rajasekar, Miguel A.F. Sanju\'an, Nonlinear Resonances, Springer, 2016.
\bibitem{Ref49}
S. Ghosh, D.S. Ray, Nonlinear vibrational resonance, Phys. Rev. E. 88(4) (2013) 042904.
\bibitem{Ref50}
J.H. Yang, Miguel A.F. Sanju\'an, H.G. Liu, Vibrational subharmonic and superharmonic resonances, Commun. Nonlinear Sci. Numer. Simulat. 30(1-3) (2016) 362-372.
\bibitem{Ref51}
D. Das, D.S. Ray, Enhancement of nonlinear response using vibrational resonance in a nonlinear oscillator; sum and difference frequency generation, Eur. Phys. J. B 91(11) (2018) 279.
\bibitem{Ref52}
L. Lou, J. Yang, Z. Wang, K. Ma, T. Gong, P. Chen, On the sampling frequency characteristics of the passive radio frequency vibration sensing tags in electromagnetic interference environments, Phys. Scr. 100(7) (2025) 075532.
\bibitem{Ref53}
L. Lou, J. Yang, K. Ma, T. Gong, Z. Wang, B. Li, A novel testing method for ultra-low-frequency vibration signal based on passive radio frequency tag sensing, Rev. Sci. Instrum. 95(9) (2024) 094702.
\bibitem{Ref54}
L. Lou, J. Yang, T. Gong, K. Ma, H. Yu, F. Tian, L. Zhao, Vibration detection under complete occlusion using passive radio frequency tag with obstructed signal paths, J. Vib. Eng. (2025), online, doi: 10.16385/j.cnki.issn.1004-4523.202505054, in Chinese.
\bibitem{Ref55}
Q. Liu, J. Yang, K. Zhang, An improved empirical wavelet transform and sensitive components selecting method for bearing fault, Measurement 187 (2022) 110348.
\bibitem{Ref56}
A. Halidou, Mohamadou Y, A.A.A. Ari, E.J.G. Zacko, Review of wavelet denoising algorithms, Multimed. Tools Appl. 82(27) (2023) 41539-41569.
\bibitem{Ref57}
M. Khodarahmi, V. Maihami, A review on Kalman filter models, Arch. Comput. Method. E. 30(1) (2023) 727-747.
%
\end{thebibliography}
\end{document}